\documentclass[a4paper,USenglish,thm-restate,cleveref,autoref]{lipics-v2021}
\hideLIPIcs

\usepackage{algorithm}
\usepackage{algpseudocode}
\usepackage{todonotes}
\usepackage{bm}
\usepackage{booktabs}
\usepackage{cases}
\usepackage{xparse}
\usepackage{xargs}
\usepackage{verbatim,xspace}
\usepackage{tikz}
\usepackage{thmtools}
\usetikzlibrary{arrows.meta, decorations.pathreplacing, calc}
\usepackage{mathtools}
\usepackage[framemethod=TikZ]{mdframed}
\usepackage{framed}
\usepackage{pgfplots}
\pgfplotsset{compat=1.18}
\usepackage[normalem]{ulem}

\newcommand{\citet}[1]{\cite{#1}}

\renewcommand{\leq}{\leqslant}
\renewcommand{\geq}{\geqslant}
\renewcommand{\le}{\leqslant}
\renewcommand{\ge}{\geqslant}

\newcounter{openquestion}

\newcommand{\II}{\mathcal{I}}
\newcommand{\JJ}{\mathcal{J}}
\newcommand{\gzsz}{\Gamma_{Z}^{(\mathsf{size})}}
\newcommand{\gxzsz}{\Gamma_{X,Z}^{(\mathsf{size})}}
\newcommand{\gazsz}{\Gamma_{A,Z}^{(\mathsf{size})}}
\newcommand{\gbzsz}{\Gamma_{B,Z}^{(\mathsf{size})}}
\newcommand{\gellzsz}{\Gamma_{\ell,Z}^{(\mathsf{size})}}
\newcommand{\gellzsm}{\Gamma_{\ell,Z}^{(\mathsf{sum})}}
\newcommand{\gazsm}{\Gamma_{A,Z}^{(\mathsf{sum})}}

\newcommand{\OL}{\widetilde{\mathcal{O}}}
\newcommand{\EE}{\mathbb{E}}
\newcommand{\yes}{\textsc{yes}}
\newcommand{\no}{\textsc{no}}

\newcommand{\DP}{\mathtt{DP}}

\newcommand{\fasttcss}{\mathtt{FasterTCSS}}
\mathchardef\mhyphen="2D

\newcommand{\ignore}[1]{}
\newcommand{\Oh}{\mathcal{O}}

\newcommand{\Ot}{\widetilde{\mathcal{O}}}

\newcommand\restr[2]{{\left.\kern-\nulldelimiterspace #1 \littletaller \right|_{#2} }}

\addtocontents{toc}{\protect\setcounter{tocdepth}{2}}

\newtheorem*{maingoal*}{Main Question}
\crefname{hypothesis}{Hypothesis}{Hypotheses}
\Crefname{hypothesis}{Hypothesis}{Hypotheses}

\newcommand{\Z}{\mathbb{Z}}

\newcommand{\C}{\mathbb{C}}
\newcommand{\F}{\mathbb{F}}

\newcommand{\G}{\mathbb{G}}

 \newcommand{\ceil}[1]{ \lceil #1 \rceil }

\newcommand{\norm}[1]{||#1||}

\def\ceil#1{\left\lceil #1 \right\rceil}

\def\norm#1{\left\| #1 \right\|}

\newcommand{\TCSS}{\ensuremath{\mathsf{TCSS}}\xspace}

\newcommand{\LCSS}{\ensuremath{\mathsf{LCSS}}\xspace}
\newcommand{\ATLCSS}{\ensuremath{\mathsf{AT\text{-}LCSS}}\xspace}

\newcommand{\PSS}{\mathsf{PSS}}

\newcommand{\fail}{\mathsf{fail}}

\title{On the Largest Cardinality Subset Sum}
\titlerunning{On the Largest Cardinality Subset Sum}

\author{Deepak Bhati}
  {CISPA Helmholtz Center for Information Security, Saarbr\"ucken, Germany}
  {deepak.bhati@cispa.de}{}{}
\author{Anubhav Dhar}
  {Max Planck Institute for Informatics, Saarbr\"ucken, Germany \and
   Saarland University, Saarbr\"ucken, Germany}
  {anubhavldhar@gmail.com}{https://orcid.org/0009-0006-5922-8300}{Supported by the Deutsche Forschungsgemeinschaft (DFG, German Research
   Foundation) grant number 559177164.}
\author{Pranjal Dutta}
  {Nanyang Technological University, Singapore}
  {pranjal.dutta@ntu.edu.sg}{https://orcid.org/0000-0001-9137-9025}{Supported by Nanyang Assistant Professorship as well as by the SUG Grant (\#025774-00001), funded by Nanyang Technological
University Singapore.}
\author{Mahesh Sreekumar Rajasree}
  {CISPA Helmholtz Center for Information Security, Saarbr\"ucken, Germany}
  {srmahesh1994@gmail.com}{https://orcid.org/0000-0002-8219-8270}
  {Funded by the European Union (ERC, LACONIC, 101041207). Views and opinions
   expressed are those of the authors only and do not necessarily reflect
   those of the European Union or the European Research Council. Neither
   the European Union nor the granting authority can be held responsible
   for them.}
\author{Karol W\k{e}grzycki}
  {Max Planck Institute for Informatics, Saarbr\"ucken, Germany}
  {kwegrzyc@mpi-inf.mpg.de}{https://orcid.org/0000-0001-9746-5733}
  {Supported by the Deutsche Forschungsgemeinschaft (DFG, German Research
   Foundation) grant number 559177164.}
\authorrunning{D. Bhati, A. Dhar, P. Dutta, M. S. Rajasree, and K. W\k{e}grzycki}
\Copyright{Deepak Bhati, Anubhav Dhar, Pranjal Dutta, Mahesh Sreekumar Rajasree, and Karol W\k{e}grzycki}


\begin{document}
\nolinenumbers
\pagenumbering{roman}
\maketitle
\thispagestyle{empty}

\begin{abstract}
We study \emph{Largest Cardinality Subset Sum} (\LCSS): given $n$
nonnegative integers and a target $t$, compute a subset summing to $t$ 
with maximum cardinality. We also consider its all-targets version
(\ATLCSS), which asks for this maximum for every sum $s\in\{1,\ldots,t\}$.
The \LCSS problem sits squarely between the subset sum and the knapsack problem. 
A straightforward dynamic-programming algorithm runs in time $\mathcal{O}(nt)$
and solves \ATLCSS, and using Jin and Bringmann 
(both in STOC $2024$, independent work) $0$-$1$ knapsack
results we achieve $\OL(n + a_{\max}^2)$-time\footnote{The notation $\OL( \cdot )$ hides polylogarithmic factors in the argument} deterministic algorithm for \LCSS.

Our main result is a randomized $\OL(n+\sqrt{n}\,t)$-time algorithm for \LCSS.
Then, we complement this result with a randomized $\OL(n+t^{3/2})$-time algorithm
for \ATLCSS. Together, our two algorithms
give an $\OL(n+t\min\{\sqrt{n},\sqrt{t}\})$-time algorithm for \LCSS.

As an application of \ATLCSS, we solve the search version of
\emph{Projection Subset Sum}, introduced by Dutta and
Rajasree~(SYNSAC 2023), in $\OL(n+t^{3/2})$ time for every fixed
$\ell_p$ norm with $p\geq1$. For $\ell_1$, we obtain
$\OL(n+t)$ time, removing the unique-representation assumption
required by their algorithm. Both algorithms output an optimal
modified vector and a witnessing subset.
\end{abstract}

\clearpage
\pagenumbering{arabic}
\section{Introduction}\label{sec:intro}

The \emph{Subset Sum} problem asks whether, given nonnegative integers
$a_1,\ldots,a_n$ and a target $t$, there exists a subset of the input items 
that sums exactly to $t$. It is a basic example of an NP-complete problem that
admits a pseudopolynomial algorithm: Bellman's dynamic program takes
$\Oh(nt)$ time~\cite{Bellman57}. Subsequent work has
substantially improved the dependence on $n$ and $t$. Koiliaris and Xu
gave a deterministic $\Ot(\sqrt n\,t)$-time algorithm~\cite{KoiliarisXu19},
and Bringmann~\cite{bringmann2017near} and, later, Jin and Wu~\cite{JinWu19} 
gave randomized $\Ot(n+t)$-time algorithms. Recently, Chan derandomized 
these results and gave a deterministic $\Tilde{\mathcal{O}}(t)$-time
algorithm for Subset sum~\cite{chan2026derandomizing}. These
algorithms determine which sums in $\{0, \ldots, t\}$ are 
\emph{attainable}, i.e.~appear as a sum of some subset of input items. 
A natural next question
is what can be learned about the subsets attaining a given target.
We study the maximum size of such a subset. Items are distinguished by
their indices, even when their values are equal. 

\begin{definition}[\textsf{Largest Cardinality Subset Sum} (\LCSS)]
\label{def:prob-LSS}
Given nonnegative integers $a_1,\ldots,a_n$ and a positive target
$t$, \LCSS asks for 
$\max\bigl\{|S|: S\subseteq\{1, \ldots, n\},\ \sum_{i\in S}a_i = t\bigr\}$,
else report if $t$ is unattainable.
\end{definition}

Thus \LCSS refines the usual decision problem: it distinguishes two
attainable representations of the same target by their cardinalities.
It also has a direct interpretation as a structured $0$-$1$ knapsack
problem: we give item $i$ weight $a_i$ and profit $Ma_i+1$, where
$M>n$, and set the knapsack capacity to $t$. Every subset of weight
exactly $t$ has profit $Mt+|S|$, whereas every subset of weight at
most $t-1$ has profit at most $M(t-1)+n<Mt$. Consequently, if $t$ is
attainable, an optimal knapsack solution is a subset that sums to $t$
and has maximum cardinality among all the subsets that sum to $t$; 
if it is unattainable, the
optimal profit is below $Mt$. This embedding explains why recording
only attainable sums, as in ordinary subset sum, does not immediately
give the desired optimum.
In fact, at the fine-grained level, this places \LCSS in between the subset sum 
and the knapsack problem; \LCSS is at least as hard as subset sum, and knapsack
is at least as hard as \LCSS. These observations make \LCSS an interesting problem to
study. 

$0$-$1$ knapsack for single capacity: 
Jin~\cite{jin20240} gave a deterministic algorithm
that runs in time $\Tilde{\mathcal{O}}(n + a_{\max}^2)$, whereas
Bringman~\cite{bringmann2024knapsack} also 
(independently) gave a deterministic $\Tilde{\mathcal{O}}(n + a_{\max}^2)$-time algorithm
for the bounded knapsack
problem (where each item comes with a multiplicity). 
Here $a_{\max}$ denotes the maximum weight item.
Using the above reduction and these two algorithms, we get 
$\Tilde{\mathcal{O}}(n + a_{\max}^2)$-time algorithm for \LCSS.

There is also a useful analogy with \emph{change-making}. The
change-making problem minimizes the number of coins summing to a
target, with unlimited use of each denomination. In \LCSS, each
input item can be used at most once, and the number of selected items
is maximized. Both problems invite a distinction between a single
specified target and all targets up to a bound. This distinction has
proved consequential for change-making: Chan and He obtained
near-linear-in-$t$ algorithms for the single-target problem and an
$\Ot(t^{4/3})$-time algorithm for its all-targets
version~\cite{ChanHe20}. We likewise study both versions of \LCSS.

\begin{definition}[\textsf{All-Targets Largest Cardinality Subset Sum}
(\ATLCSS)]\label{def:prob-AT-LCSS}
Given nonnegative integers $a_1,\ldots,a_n$ and a positive integer $t$,
\ATLCSS asks, for every $s\in\{1,\ldots,t\}$, for the maximum
cardinality of a subset $S\subseteq\{1, \ldots, n\}$ with $\Sigma(S)=s$, 
else report if $s$ is unattainable.
\end{definition}

A straightforward dynamic program computes the optimum for every sum up to $t$
in $\Oh(nt)$ time: when item $a_i$ is processed, the best cardinality at
sum $s$ is compared with one plus the previous best cardinality at
$s-a_i$. Koiliaris and Xu~\cite{KoiliarisXu19} also discuss the problem of subset sums with cardinality 
information. They further provide an $\Tilde{\mathcal{O}}(nt)$-time algorithm that
gives all the attainable subset sums with complete cardinality information, 
not just the maximum. However, their improvement of the complexity from 
$\Tilde{\mathcal{O}}(nt)$ to $\Tilde{\mathcal{O}}(\sqrt{n}t)$ loses
information regarding cardinalities. This suggests that solving \LCSS 
in $\Tilde{\mathcal{O}}(\sqrt{n}t)$ time is an interesting problem. 
On the other hand, Axiotis and Tzamos~\cite{axiotis2018capacitated} 
give an $\mathcal{O}(Dt)$-time algorithm for the knapsack problem,
with all weights being less than or equal to $t$, where $D$ denotes the
number of distinct items in the instance. This algorithm, with the aforementioned reduction
of \LCSS to knapsack, gives a $\mathcal{O}(Dt)$-time algorithm
for \ATLCSS, and consequently for \LCSS too.

The subset-sum algorithms in \cite{bringmann2017near,JinWu19,chan2026derandomizing} do
not, on their own, preserve the largest cardinality for each sum. 
Related work has examined other forms of cardinality information:
Dutta and Rajasree~\cite{DuttaRajasree22} study the Hamming weights 
of all solutions when the number of solutions is bounded, 
while Antonopoulos, Pagourtzis, Petsalakis, and Vasilakis~\cite{AntonopoulosEtAl23}
discuss algorithms for subset-sum variants with target-cardinality
constraints.

\subsection{Projection Subset Sum and related work}\label{sec:related-work}

\ATLCSS has a direct application to
\emph{Projection Subset Sum}, introduced by Dutta and
Rajasree~\cite{DR23}. Given a nonnegative integer vector
$Z=(a_1,\ldots,a_n)$ and a nonnegative integer target $t$, the
search problem asks for an integer vector $X=(x_1,\ldots,x_n)$ and a
subset $S\subseteq[n]$ satisfying $\sum_{i\in S}x_i=t$ that minimize
$\norm{Z-X}_p$. For $t>0$, the set $S$ is necessarily nonempty.
Their near-linear randomized
algorithm for the $\ell_1$ case assumes that every attainable sum has
a unique representation; they also give an $\Oh(nt)$ algorithm under
that assumption~\cite{DR23}. For a fixed nonempty subset $S$, the
amount by which its original sum misses $t$ is
$d=t-\Sigma(S)$. In the $\ell_\infty$ case, if integer-valued
coordinates may be changed without sign restrictions, the smallest
possible maximum coordinate change is $\left\lceil{|d|}/{|S|}\right\rceil$.
More selected items allow the discrepancy to be spread over more
coordinates. For general $p\geq1$, the corresponding exact integer
cost further depends on $\Sigma(S)$ and $|S|$; see
\Cref{sec:PSS}. Among nonempty subsets with the same sum, a largest
subset minimizes the projection cost. Therefore, an \ATLCSS solution supplies exactly the cardinality information
needed to compare the best projection costs for different attainable
sums, including sums above $t$.

A related motivation for Projection Subset Sum comes from neural-network pruning. The strong lottery ticket hypothesis asks whether a randomly initialized network contains a subnetwork that approximates a given target network~\cite{ramanujan2020s,malach2020proving}. Pensia et al.~\cite{PensiaEtAl20} study this question using random subset sums, while Xiong, Liao, and Kyrillidis~\cite{XiongLiaoKyrillidis23} additionally allow bounded perturbations of the weights. This combines selecting existing weights with slightly modifying them to improve approximation. Projection Subset Sum captures a related question for a list of numbers: how little must the numbers be changed so that some subset sums exactly to a prescribed target?

\subsection{Our results}

The previous discussion 
already gives us $\OL(n + a_{\max}^2)$-time and $\mathcal{O}(n + Dt)$-time algorithms for \LCSS. Our main result is a 
faster pseudopolynomial-time algorithm for the regime when 
$a_{\max} = \widetilde\omega(n^{1/4}\sqrt{t})$ and $D = \widetilde\omega(\sqrt{n})$. Note that $\widetilde\omega(\cdot)$ hides polylogarithmic factors.

\begin{restatable}{theorem}{fasLSSalgo}\label{thm:fas-LSS-algo}
For any fixed integer constant $c\geq 10$, there exists a randomized
algorithm that solves \LCSS on $n$ nonnegative integers with target
$t$ in time $\OL(n+\sqrt{n}\,t)$, with error probability at most
$n^{-c}$.
\end{restatable}

Thus, up to polylogarithmic factors, our algorithm reduces the
coefficient of $t$ from $n$ to $\sqrt{n}$ in the running time complexity. It computes the exact maximum cardinality of a subset summing to $t$, 
or reports that $t$ is unattainable, with high probability.

Very recently, Jin and Mao~\cite[Proposition~D.3]{JinMao26} give a \emph{conditional result}
for \LCSS on positive inputs.  If there is an algorithm running in $T(n,\Delta)$
that decomposes all subset sums of $n$ positive integers bounded by $\Delta$
into an arithmetic progression with a common, then they managed to show that \LCSS
can be solved in $\Ot(n+\sqrt{n}\cdot t + T(n,3t\sqrt{n}))$ time.  Their result
would match the running time of our result, if $T(n,\Delta) \le \Ot(n +
\Delta)$, however the existance of such a fast algorithm is
open~\cite{JinMao26}.  \Cref{thm:fas-LSS-algo} avoids this unresolved assumption
and achieves the same running-time bound \emph{unconditionally}.

Our second result focuses on the all-targets version, in which
the maximum cardinality must be computed for every target
$s\in\{1,\ldots,t\}$.

\begin{restatable}{theorem}{allLSSalgo}\label{thm:all-LSS-algo}
For any fixed integer constant $c\geq 10$, there exists a randomized
algorithm that solves \ATLCSS on $n$ nonnegative integers with target $t$ in time
$\OL(n+t^{3/2})$, with failure probability at most $n^{-c}$.
\end{restatable}

This algorithm improves the discussed $\Oh(n + Dt)$ algorithm when $D = \widetilde\omega\left(t^{1/2}\right)$.
The probability guarantee holds simultaneously for all targets.
Beyond the time needed to read the input, the running time depends
only on $t$, rather than on the number of items. 
The two bounds are complementary. Since the all-targets algorithm
also solves any single-target instance, together they yield an
$\Ot(n+t\min\{\sqrt{n},\sqrt{t}\})$-time algorithm for \LCSS.

By exploiting the cardinality information provided by \ATLCSS, 
we also give a randomized $\OL(n+t^{3/2})$-time algorithm
for Projection Subset Sum in the $\ell_p$ norm ($\PSS_p$) for all fixed $p\geq1$.
\begin{restatable}{theorem}{pssallnormsalgo}\label{thm:pss-all-norms-algo} 
    For fixed $p\geq1$, given nonnegative integers $Z=(a_1,\ldots,a_n)$
    and an integer $t\geq0$, there is a randomized algorithm that outputs
    an optimal solution $(X,S)$ to $\PSS_p$ in time $\OL(n+t^{3/2})$,
    with failure probability at most $1/3$.
\end{restatable}
And, for the $\ell_1$ norm, we give
a randomized $\OL(n+t)$-time algorithm using Bringmann's subset-sum
algorithm; see \Cref{thm:pss-l1-algo}. Both the $\PSS$ algorithms
output an optimal vector $X$ and a set $S$ witnessing its feasibility,
with failure probability at most $1/3$, without the unique-representation
assumption.

\subparagraph*{Organization of the paper.}
\Cref{sec:overview} gives an overview of our algorithms.
\cref{sec:preliminaries} introduces the notation and tools used. \Cref{sec:single-target} presents the algorithm for \LCSS via solving \TCSS. The algorithm for \ATLCSS
can be found in \cref{sec:all-targets}. The algorithms for Projection Subset Sum can be found in 
\cref{sec:PSS}. \cref{sec:deferred-proofs}
contains deferred lemmas and proofs.

\subparagraph*{Declaration on the Use of AI Tools.}
Generative AI tools, including ChatGPT and Claude, were used solely for proofreading, language editing, improving presentation, and identifying potential inconsistencies. All technical contributions are original to the authors, who independently verified relevant AI-generated suggestions and take full responsibility for the manuscript. \section{Technical Overview}\label{sec:overview}

Previously, Bringmann~\cite{bringmann2017near}
gave an $\OL(n + t)$-time algorithm to output all possible subset sums capped by 
an integer $t$, in a collection of $n$ elements. This was later simplified by Jin 
and Wu~\cite{JinWu19}, and then derandomized by Chan~\cite{chan2026derandomizing}. 
For this paper, we give a bivariate equivalent result of~\cite{JinWu19}, 
which computes both cardinalities and sums of all possible
subsets with caps in both cardinality and the sum.
\begin{restatable}{lemma}{AllTCardBound}\label{lem:all-targets-bounded-cardinality} 
    Given nonnegative integers $a_1,\ldots,a_n$, positive integers
    $k$ and $t$, and a parameter $\delta \in (0, 1)$, there is a
    randomized algorithm with one-sided error probability at most 
    $\delta$ that computes the set 
    $\{(\Sigma(S),|S|):S\subseteq\{1, \ldots, n\},\ |S|\le k,\ \Sigma(S)\le t\}$
    in time $$\Oh((n+kt)\operatorname{polylog}(n + kt + 1/\delta)).$$
    The returned set is always a subset of 
    $\{(\Sigma(S),|S|):S\subseteq\{1, \ldots, n\},\ |S|\le k,\ \Sigma(S)\le t\}$.
\end{restatable}
A proof idea is as follows: Consider Bringmann's algorithm~\cite{bringmann2017near}.
Essentially, the algorithm consists of a collection of convolutions to 
keep track of sums. Whenever a sum exceeds $t$, this sum is dropped from further
consideration. One can modify Bringmann's algorithm to simply consider pairs of 
cardinalities and sums, then convolve these sets of pairs, and drop pairs which
exceed either the cardinality bound or the sum cap. The resulting algorithm has
the required running time. Similarly, Jin and Wu's algorithm~\cite{JinWu19} can 
also be modified accordingly to prove \cref{lem:all-targets-bounded-cardinality};
for completeness, we provide a complete proof in 
\cref{sec:proof-of-all-targets-bounded-cardinality}.

\subsection{Solving Largest Cardinality Subset Sum}

We start by focusing on the intermediate problem of finding whether a subset of a given sum
and a given cardinality exists.

\begin{definition}[\textsf{Target Cardinality Subset Sum} (\TCSS)]
\label{def:prob-TCSS}
Let $a_1,\ldots,a_n,k,t$ be nonnegative integers. \TCSS asks whether
there is a subset
$S\subseteq\{1, \ldots, n\}$ with $|S|=k$ and $\Sigma(S)=t$.
\end{definition}
We first provide an $\OL(n + \sqrt{n}t)$ algorithm for \TCSS, and then 
use it to solve \LCSS. 

\subparagraph*{Algorithm for \TCSS.}
Our approach can be informally described as follows. We process items smaller than $t/\sqrt{n}$ and items larger than $t/\sqrt{n}$ separately. At most $\sqrt{n}$
many large items can contribute to a solution; we use \cref{lem:all-targets-bounded-cardinality} to essentially list all cardinalities and sums of all relevant subsets with at most $\sqrt{n}$ large items.
Next, we color the indices
randomly into $\sqrt{n}$ colors. For each color class, we compute all 
cardinality--sum pairs of the small items (using~\cref{lem:all-targets-bounded-cardinality}) with a sum cap of $\OL(t/\sqrt{n})$.
We merge these cardinality--sum pairs in a merge-sort fashion: at each 
step, merge two sets of cardinality--sum pairs to get a larger one. However, doing this
naively would blow up the sizes of these sets. We overcome this issue by keeping
only the pairs whose cardinality deviates from the expected cardinality of any 
solution by at most $\OL(\sqrt{n})$, while discarding others. This crucial step 
aids in making the algorithm efficient ($\OL(\sqrt{n} t)$ time spent on each level), 
while the Chernoff bound guarantees that with high probability, a solution is not discarded. Finally, our algorithm decides if a solution exists by checking the existence of a compatible subset among the small items for each cardinality--sum pair of the large items.
An overview of the details and the techniques is as follows. 

Assume that the number of items $n$ is a power of $4$, and all items are bounded 
by $\Theta(t)$. We start by color coding 
the indices into $\sqrt{n}$ color classes. Formally, construct a random partition
$Z_1 \uplus \cdots \uplus Z_{\sqrt{n}} = \{1 , \ldots, n\}$ of the indices. We
define $Z_j^{(\ell)} = Z_{2^\ell (j - 1) + 1} \cup Z_{2^\ell (j - 1) + 2} \cup 
\cdots \cup Z_{2^\ell (j - 1) + 2^{\ell}}$, for $\ell \le \frac{1}{2}\log n$.

Assume that a solution exists, i.e.~let $X \subseteq \{1, \ldots, n\}$ be such 
that $\Sigma(X) = t$ and $|X| = k$. Consider the following breakup of the input
elements into heavy elements $B$ and light elements $A$, with their respective 
intersection with the solutions being $X_B$ and $X_A$.
\begin{align*}
B &\coloneq \left\{i\in\{1, \ldots, n\} : a_i\in\left(\frac{t}{\sqrt n},t+1\right]\right\},
& X_B &\coloneq X\cap B,\\
A &\coloneq \{1, \ldots, n\} \setminus B,
& X_A &\coloneq X\cap A.
\end{align*}
It is easy to see that $X_B$ has size at most $\sqrt{n}$, as all elements of 
$B$ have a value of at least $t / \sqrt{n}$. Using 
\cref{lem:all-targets-bounded-cardinality} on the indices in $B$, we compute
\begin{equation*}\label{eq:jj-to}
    \JJ \gets \{(|S|, \Sigma(S)) : S \subseteq B, |S| \le \sqrt{n}, \Sigma(S) \le t\}
\end{equation*}
in time $\OL(n + \sqrt{n}t)$. We must have $(|X_B|, \Sigma(X_B)) \in \JJ$.

We now focus on the light items. Using the Chernoff bound, we show that (i) each
color class is of size $\Oh(\sqrt{n})$, (ii) the size of intersection of $X_A$ with a given
subset of color classes deviates by at most $\OL(\sqrt{n})$ from its expected
intersection size, and (iii) the sum of elements in $X_A$ from any fixed color class is at
most $\OL(t/\sqrt{n})$, i.e.~the color classes split the sum somewhat evenly. More
formally, we prove the following lemma.

\newtheorem*{simplifiedlemma}{\cref{lem:desired-whp-property} (Simplified)}
\begin{simplifiedlemma}
With probability at least $1-n^{-3c}$, we have all of the following.
\begin{enumerate}
    \item $|Z_j|\leq\Oh(\sqrt n)$, for $j\in\{1, \ldots, \sqrt n\}$,
    \item $\left||X_A\cap Z_j^{(\ell)}|-\frac{2^\ell k}{\sqrt n}\right|\le\Oh(\sqrt n\log n)$, for $\ell\in\{0, 1, \ldots, \frac12\log n\}$, $j\in\{1, \ldots, \sqrt n/2^\ell\}$, and
    \item $\Sigma(X_A\cap Z_j)\leq\Oh((t/\sqrt n)\log^2 n)$, for $j\in\{1, \ldots, \sqrt n\}$.
\end{enumerate}
\end{simplifiedlemma}

We now use these guarantees to compute the cardinalities and sums of relevant subsets
of the light items. For each $j\in\{1,\ldots,\sqrt n\}$, we use
\cref{lem:all-targets-bounded-cardinality} on $(a_i)_{i\in Z_j\cap A}$,
with cardinality bound $\Oh(\sqrt{n})$ and a cap of
$\Oh((t/\sqrt n)\log^2 n)$ on the sum,
to compute
\begin{align*}
 \II_j^{(0)}=\biggl\{\left(|S|-\left\lfloor\frac{k}{\sqrt n}\right\rfloor,
                         \Sigma(S)\right):\;&S\subseteq Z_j\cap A, \quad \Sigma(S)\leq\Oh((t/\sqrt n)\log^2 n),\\
 &\left||S|-\left\lfloor\frac{k}{\sqrt n}\right\rfloor\right|
       \leq\Oh(\sqrt n\log n)\biggr\}.
\end{align*}
An intuition behind $\II_j^{(0)}$ is as follows. Notice that the expected 
value of $|X \cap Z_j|$ over the randomness of color coding is exactly 
$\frac{k}{\sqrt{n}}$; we only consider 
$\left\lfloor \frac{k}{\sqrt{n}} \right\rfloor$ simply to work
with integers, while staying close to the actual expected value. Moreover, since $|X \setminus X_A| = |X_B| \le \sqrt{n}$, it can be shown that the expected value of $|X_A \cap Z_j|$ is in $\left(\frac{k}{\sqrt{n}} - \Oh(\log n), \frac{k}{\sqrt{n}} + \Oh(\log n)\right)$. Therefore, 
a pair $(d, w) \in \II_j^{(0)}$ corresponds to the existence of (at 
least) one subset $S \subseteq Z_j \cap A$ which sums up to $w$ and has 
deviation of $d \pm \Oh(\log n)$ from the expected cardinality of $|X_A \cap Z_j|$. 
In fact, another application of the Chernoff bound shows that a pair 
corresponding to $X_A \cap Z_j$ is in the set $\II_j^{(0)}$ with high 
probability. Note that computing $\II^{(0)}_j$ for all color classes takes 
time $\OL(n+\sqrt n\,t)$.

Continuing in the same spirit of working with pairs representing (i) deviation 
from expected size, and (ii) sum of the subset, our next aim is to obtain such pairs for
subsets $S \subseteq Z^{(\ell)}_j \cap A$, even for $\ell > 0$. 
Observe that $Z^{(\ell)}_j = Z_{2j - 1}^{(\ell - 1)} \cup  Z_{2j}^{(\ell - 1)}$. If a subset $S_1 \subseteq Z^{(\ell - 1)}_{2j - 1}$ deviates from the expected size $|X \cap Z_{2j - 1}^{(\ell - 1)}|$ (i.e.~$\frac{2^{\ell - 1}k}{\sqrt{n}}$) by $d_1'$, and a subset $S_2 \subseteq Z^{(\ell - 1)}_{2j}$ deviates from the expected size $|X \cap Z_{2j}^{(\ell - 1)}|$ (which is again $\frac{2^{\ell - 1}k}{\sqrt{n}}$) by $d'_2$, then their union $S_1 \cup S_2$ deviates from the expected size of $|X \cap Z_{j}^{(\ell)}|$ by $d_1' + d'_2$. Moreover, the sum of $S_1 \cup S_2$ is the sum of $S_1$ plus the sum of $S_2$.
This allows us to define a set $\II^{(\ell)}_j$ of pairs and compute them recursively by taking the sum sets of the two lower-level sets $\II^{(\ell - 1)}_{2j - 1}$ and $\II^{(\ell - 1)}_{2j}$. Since we are working with the floor of the expectation, we must be slightly careful in the recursion computation. Formally,
for $1 \le \ell \le \frac{1}{2}\log n$ and $1 \le j \le \sqrt{n} / 2^\ell$,
we recursively compute
\begin{align*}
 \II_j^{(\ell)}=\biggl\{(d,w&):\;
 (d_1,w_1)\in\II_{2j-1}^{(\ell-1)}, \quad (d_2, w_2) \in \II_{2j}^{(\ell-1)}, \quad w = w_1 + w_2\\
 &d= \left(d_1 + \left\lfloor\frac{2^{\ell-1}k}{\sqrt n}\right\rfloor \right) + \left(d_2 + \left\lfloor\frac{2^{\ell-1}k}{\sqrt n}\right\rfloor
 \right)
       -\left\lfloor\frac{2^\ell k}{\sqrt n}\right\rfloor,
 \quad |d|\leq\Oh(\sqrt n\log n)\biggr\}.
\end{align*}

Note that we further trim the sets with the condition 
$|d|\leq\Oh(\sqrt n\log n)$. This ensures that the sizes of these sets
are bounded by $\OL(2^{\ell} \cdot t + \sqrt{n})$. Further, it is enough to 
guarantee (from the Chernoff bound) that with high probability, 
the pair corresponding to deviation and sum for $X_A \cap Z^{(\ell)}_j$
survives this trimming,
(in this and all other lower levels of the recursion) and stays in the 
set $\II_j^{(\ell)}$. In particular, with high probability, the pair
$(|X_A| - k, \Sigma(X_A)) = (-|X_B|,t-\Sigma(X_B))$ is in the set 
$\II^{(\frac{1}{2}\log n)}_1$. We therefore return \yes\ if some
$(s,w)\in\JJ$ satisfies $(-s,t-w)\in\II_1^{(\frac12\log n)}$,
and \no\ otherwise. Such a pair gives disjoint subsets of $B$ and $A$
whose union has cardinality $k$ and sum $t$.

It remains to bound the time for computing the sets 
$\II^{(\ell)}_j$. Every $(d,w)\in\II_j^{(\ell)}$ satisfies
$|d|\leq\Oh(\sqrt n\log n)$ and it can be shown that 
$w\leq\OL(2^\ell t/\sqrt n + 1)$. Hence computing $\II^{(\ell)}_j$ takes time $\OL\bigl(\sqrt n \cdot( 2^{\ell-1}t/\sqrt n + 1)\bigr)=\OL(2^\ell t + \sqrt{n})$ using a fast convolution algorithm (see \cref{lem:pair-sumset}).
There are $\sqrt n/2^\ell$ such computations for a fixed value of $\ell$, and $\frac{1}{2}\log n$ many values of $\ell$. Hence, the total running time amounts to
$
 \sum_{\ell=1}^{\frac12\log n}\frac{\sqrt n}{2^\ell}
       \OL(2^\ell t + \sqrt{n})=\OL(n  + \sqrt n\,t)
$.
Together with the other steps, the total running time is 
$\OL(n+\sqrt n\,t)$. The details can be found in \cref{thm:fast-tcss}.

\subparagraph*{From \TCSS to \LCSS.}

To solve \LCSS, we first decide whether there exists a subset with sum $t$
and cardinality in an interval $[\ell,h]$: This can be done by appending
$h-\ell$ zero-valued items and calling a \TCSS algorithm with target sum 
$t$ and target cardinality $h$. With this, we can binary search for the 
largest cardinality of a subset summing up to $t$.

We first check if $t$ is a possible subset sum; if not, we terminate.
Henceforth, assume that there exists a subset summing up to $t$. Let 
$k^*$ be the largest cardinality of a subset summing to $t$. We use 
binary search, starting with $\ell=0$ and $h=n+1$, and maintain the 
invariant of $k^*\in[\ell,h)$ throughout. At each iteration,
we set $m=\lfloor(\ell+h)/2\rfloor$, and check if there exists a subset
with sum $t$ and cardinality in $[m,h)$. Following the binary search
framework, we set $\ell=m$ if such a subset exists; otherwise we
set $h=m$. When $h=\ell+1$, we output $\ell=k^*$. 

There are $\Oh(\log n)$ calls, each on $\Oh(n)$ items, giving a total
running time of $\OL(n+\sqrt n\,t)$. More details can be found in 
\cref{sec:single-target}.

It still remains open whether one can get better algorithms for \LCSS, possibly with a running time of $\OL(n + t)$.
\subsection{Solving All-targets Largest Cardinality Subset Sum}

For \ATLCSS, we assume without loss of generality that all items are positive 
(otherwise we simply delete all zero items and add back their count in all outputs).
Bellman's dynamic program for knapsack~\cite{GareyJohnson79,Bellman57} can be adapted to give us an $\Oh(nt)$ algorithm for \ATLCSS. We provide an algorithm with running time $\OL(n + t^{3/2})$, which is an improvement 
in the regime $t \le n^{2 - \varepsilon}$, for some $\varepsilon > 0$.

We follow the framework by Bringmann and Kociumaka as discussed in 
\cite{lincoln2020monochromatic} (also see \cite[Section 3]{ChanHe20}). They consider
the all-targets change-making problem; recall that given a set of denominations, 
the problem asks for the smallest number of coins which sum up to $1$, $2$, \ldots, 
$t$ (as discussed in \cref{sec:intro}). The framework can be broadly described as follows.
\begin{enumerate}
    \item Set a threshold $\tau$. Classify denominations larger than 
    $\tau$ as \emph{heavy} and the others \emph{light}.
    \item Compute relevant information from the heavy denominations in 
    time $\OL(t^2/\tau)$. The crucial observation is to notice that in any solution, 
    at most $t/\tau$ many coins appear from a heavy denomination.
    \item Incorporate on top all light denominations using a classical dynamic 
    programming approach, which takes a total time of $\OL(t\tau)$. The 
    crucial observation here is to notice that there are at most $\tau$ many light 
    denominations.
\end{enumerate}
This framework produces an algorithm with a running-time of 
$\OL(n + t^2/\tau + t\tau)$. Setting $\tau = \Theta(\sqrt{t})$ minimizes this to 
$\OL(n + t^{3/2})$.

Although \ATLCSS can be viewed as a $0$-$1$ variant of the all-targets change-making 
problem, the specific algorithm for change-making does not directly give us an
algorithm for \ATLCSS. We follow this broad framework of splitting into heavy and 
light items, but modify the specifics as follows.

\subparagraph*{Handling heavy items.} 
Firstly, the algorithm for handling heavy denominations does not directly work for 
\ATLCSS. We overcome this obstacle using \cref{lem:all-targets-bounded-cardinality} 
instead: We compute the cardinalities and sums of all subsets of the heavy items with 
sum at most $t$ and cardinality at most $t / \tau$. This takes a total time of 
$\OL(n + t^2/\tau)$ and allows us to extract the necessary information regarding the
heavy items.

\subparagraph*{Reducing the count of light items.}
Secondly, the algorithm for light denomination crucially requires the number of light 
denominations to be bounded by $\tau$. This is not the case for \ATLCSS; in fact, it
is quite possible that all input items in an \ATLCSS are light, for some $\tau$.
Moreover, it is important to keep track of the exact multiplicities of items. We 
resolve this issue using a multiplicity reduction trick by 
Lawler~\cite{lawler1979fast}  (see also the discussion on preprocessing multisets in 
\cite[Section 2]{bringmann2017near}) as follows. Consider each item initially having 
a \emph{profit} of $1$, then maximizing cardinality is equivalent to maximizing 
profit. Whenever there are at least three copies of items with value $w$ and profit 
$p$, we delete two such copies and add a new element with value $2w$ and profit $2p$. 
Such an operation is safe as the maximum profits for all target sums remain
unchanged. Moreover, this additionally guarantees the number of items is at most 
$\Oh(\tau \log t)$. Moreover, this can be done in time $\Oh(n + \tau \log t)$.

\subparagraph*{Solving All-Targets Largest Cardinality Subset Sum.}
We present an overview of the algorithm solving \ATLCSS following the above framework.
First, we count the number of zeros $z$ appearing in our input, and delete all such 
items. We add back $z$ to every output at the final step of our algorithm. Now, fix a 
threshold $\tau \in \{1, \ldots, t\}$. Let 
$H = \{i \in \{1, \ldots, n\} : a_i > \tau\}$ be the indices of the heavy elements, 
and let $L = \{i \in \{1, \ldots, n\} : a_i \le \tau\}$ be the indices of the light 
elements.

We start by computing the output of \ATLCSS for just the heavy items. For
$s \in \{0, \ldots, t\}$, let $\mathtt{largest}(s)$ denote the largest cardinality of 
a subset of the heavy items summing up to exactly $s$. Since all heavy items have 
value at least $\tau$, we must have $\mathtt{largest}(s) \le s / \tau \le t / \tau$. 
Therefore to compute $\mathtt{largest}(s)$ for all $s \in \{0, \ldots, t\}$, we use
\cref{lem:all-targets-bounded-cardinality} on $(a_i)_{i \in H}$ with cardinality
bound $t / \tau$ and a cap of $t$ on the sum. That is, we compute
\[\{(\Sigma(S), |S|) : S \subseteq H, |S| \le t / \tau, \Sigma(S) \le t\};\]
this can be done in time $\OL(n + t^2/\tau)$. Then we iterate over all pairs
$(\Sigma(S), |S|)$, and update $\mathtt{largest}(\Sigma(S))$ to $|S|$, if this causes
an increase in its value.

To incorporate the light items, we use the multiplicity reduction trick by Lawler,
i.e.~in time $\Oh(n + \tau\log t)$, we transform the light items into a multiset $\mathcal L$ of at most  $\Oh(\tau \log t)$ pairs of value and profit.

This allows us to use a dynamic programming approach as follows. First, let $\DP_j[s]$ 
denote the maximum of the number of some selected heavy items plus the profit of 
some selected pairs among the first $j$ pairs of $\mathcal L$, such that their 
total weight is $s$. Formally, if $\mathcal L$ contains the elements $(w_1, p_1)$, 
\ldots, $(w_m, p_m)$, where $w_1, \ldots, w_m$ are the values and $p_1, \ldots, p_m$ 
are the respective profits, then for $0\leq j\leq m$ and $0\leq s\leq t$,
\[
 \DP_j[s]:=
 \max\left\{|S|+\sum\nolimits_{i\in I}p_i :
 S\subseteq H,\ I\subseteq\{1, \ldots, j\},
 \Sigma(S)+\sum\nolimits_{i\in I}w_i=s\right\}.
\]
For the base case, we set $\DP_0[s] = \mathtt{largest}(s)$ for all 
$s \in \{0, \ldots, t\}$; these are the largest cardinalities of subsets summing up 
to $s$, when no elements from $\mathcal L$ are considered. For transitions, notice 
that $\DP_j[s]$ can either contain the $j$-th pair or not contain it. We take the 
maximum of the two. Formally, 
\[
 \DP_j[s]=
 \begin{cases}
  \DP_{j-1}[s],&s<w_j,\\
  \max\{\DP_{j-1}[s],\ \DP_{j-1}[s-w_j]+p_j\},&s\geq w_j
 \end{cases}.
\]
After processing all items in $\mathcal L$, the answer is stored in $\DP_m[s]$ for
all $s \in \{0, \ldots, t\}$. We finally add back the count of the deleted 
zero-valued items giving us the output
\[(\DP_m[0] + z), (\DP_m[1] + z), \ldots, (\DP_m[t] + z).\]
The dynamic program takes time $\Oh(|\mathcal L| \cdot t) = \OL(t \tau)$.
The total running time of the algorithm is therefore $\OL(n + t^2/\tau + t\tau)$.
Selecting $\tau = \Theta(\sqrt{t})$ minimizes this running time to 
$\OL(n + t^{3/2})$. The details can be found in \cref{sec:all-targets}.

It is interesting to note that although it is possible to modify Bringmann and 
Kociumaka's framework for \ATLCSS, the improvements by Chan and He~\cite{ChanHe20} 
do not extend directly. For the all targets change-making problem, they handle the
heavy items identically, however for incorporating the light items a 
faster approach using binary $(\min,+)$-convolution is used, making this step take 
merely $\OL(t\sqrt{\tau})$ time. Their algorithm hence has a total running time of 
$\OL(n + t^2 / \tau + t\sqrt{\tau})$. This is minimized by setting 
$\tau = \Theta(t^{2/3})$ to $\OL(n + t^{4/3})$. This approach crucially uses
the fact that coins of a particular denomination can be selected an arbitrary number of
times as opposed to the $0$-$1$ nature of \ATLCSS. It remains open to get 
$\OL(n + t^{4/3})$-time algorithms for \ATLCSS, possibly using a smarter 
modification of the framework by Chan and He.
 \section{Preliminaries}
\label{sec:preliminaries}

All logarithms are base $2$. For a positive integer $m$, we define $[m] := \{1, \ldots, m\}$ and $[m]_0 := \{0, \ldots, m\}$. 
We consider a multiset $Z$ containing $n$ non-negative integers $a_1, a_2, \ldots, a_n$. We denote 
$\Sigma(Z) = \sum_{i \in Z} a_i$. For sets $P,Q\subseteq\mathbb Z^2$ of pairs, we define their convolution $P\oplus Q$ as $\{(x_1+x_2,y_1+y_2):(x_1,y_1)\in P,\ (x_2,y_2)\in Q\}$. This can be computed efficiently (see \cref{lem:pair-sumset} in \cref{sec:pair-sumset}).
For completeness, we also state the well-known Chernoff bound.

\begin{lemma}[Chernoff Bound]\label{lem:chernoff}
    Let $Y_1, Y_2, \ldots, Y_n$ be independent Bernoulli random variables. Let $Y = Y_1 + Y_2 + \cdots + Y_n$ and let $\mu = \EE[Y]$. Then, for all $\delta \in (0,1)$,
    \[\Pr\left[|Y - \mu| \ge \delta \mu \right] \le 2\exp\left(- \frac{\delta^2 \mu}{3}\right).\]
    Moreover, for any $\lambda \ge 2e\mu$, we have $\Pr[|Y| \ge \lambda] \le 2^{-\lambda}$.
\end{lemma}
 \section{Largest Cardinality Subset Sum}  \label{sec:single-target}

In this section we present a randomized algorithm for \LCSS on nonnegative
integers $a_1,\ldots,a_n$ and target $t$ (recall \cref{def:prob-LSS}). We first describe an algorithm for
\TCSS with target cardinality $k$ (recall \cref{def:prob-TCSS}), and then use it as a subroutine to solve \LCSS. 
The running time of both algorithms is $\OL(n+\sqrt{n}\,t)$.

\subsection{Algorithm for Target Cardinality Subset Sum}

We may assume the following:
\begin{itemize}
    \item $a_i\in[t+1]_0$ for every $i\in[n]$ (if not, we replace
          values greater than $t$ by $t+1$), and
    \item $n$ is a power of four (if not, we simply add dummy items of value $t+1$).
\end{itemize}
The last assumption makes sure that $\sqrt{n}$, $\frac{1}{2}\log n$, etc are integers.

\subsubsection{Color coding.} \label{sec:col-cod}

We perform a random color coding of the indices $[n]$ into $\sqrt{n}$ colors. Formally, 
let $Z_1 \uplus Z_2 \uplus \cdots \uplus Z_{\sqrt{n}}$ be a random partition of $[n]$. That is, for each index $i \in [n]$, an integer $j \in [\sqrt{n}]$ is chosen uniformly at random, and $i$ is added to $Z_j$. We refer to these sets $Z_j$ as \emph{color classes}. For $\ell \in [\frac{1}{2} \log n]_0$ and $j\in[\sqrt n/2^\ell]$, let $Z^{(\ell)}_j = Z_{2^\ell (j - 1) + 1} \cup Z_{2^\ell (j - 1) + 2} \cup \cdots \cup Z_{2^\ell (j - 1) + 2^{\ell}}$. 

\subsubsection{Segregating the input and the solution.} \label{sec:input-partition}
Assuming there exists a solution, we fix a solution $X\subseteq[n]$, i.e.~with $\Sigma(X)=t$ and $|X|=k$. First, we segregate the items into heavy and light weights as follows. 
\begin{align*}
B &\coloneq \left\{i\in[n]: a_i\in\left(\frac{t}{\sqrt n},t+1\right]\right\},
& X_B &\coloneq X\cap B,\\
A &\coloneq [n]\setminus B,
& X_A &\coloneq X\cap A.
\end{align*}
We also define the following sets. 
\begin{align*}
    C &\coloneq \left\{i\in[n]: a_i\in\left[0,\frac{t}{n\sqrt n}\right]\right\},
    & X_C &\coloneq X\cap C,\\
    L_\ell &\coloneq \left\{i\in[n]: a_i\in\left(\frac{t}{2^\ell},\frac{t}{2^{\ell-1}}\right]\right\},
    & X_\ell &\coloneq X\cap L_\ell.
\end{align*}
Here $L_\ell$ and $X_\ell$ are defined for every integer
$\ell\in\{\frac12\log n+1,\ldots,\frac32\log n\}$.
We refer to the sets $L_\ell$ as \emph{layers}.
We observe that these sets partition the sets $[n]$ and $X$ respectively;
\[[n] = B \uplus C \uplus \left( \bigcup \nolimits_{\ell = \frac{1}{2}\log n + 1}^{\frac{3}{2} \log n} L_\ell \right), \quad \text{and} \quad X = X_B \uplus X_C \uplus \left( \bigcup\nolimits_{\ell = \frac{1}{2}\log n + 1}^{\frac{3}{2} \log n} X_\ell \right).\]
These sets will help us define and derive various probabilistic guarantees. 
These, in turn, culminate in useful structural properties that we leverage to design our algorithm.

\subsubsection{Probabilistic guarantees.}\label{sec:prob-guarantee}
The goal of this section is to prove that the randomness in the color coding step guarantees that the input would
be somewhat well structured. We prove various guarantees and unify them later in \cref{lem:desired-whp-property}. All the probabilities in this section are over the 
random choice of the partition, i.e.~the color coding. Moreover, recall that $c \ge 10$ is a fixed constant.

We begin by showing that each color class has $\Oh(\sqrt{n})$ items with high probability.
This bounds the size of the sub-instances processed by the algorithm. For this bound, we define
\begin{equation}\label{eq:gzsz}
\gzsz\coloneq2e\sqrt n+4c\log n \le \Oh(\sqrt n),
\end{equation}

\begin{lemma}\label{lem:Zi-size-bound}
For every $j\in[\sqrt{n}]$,
$
 \Pr[|Z_j|\geq\gzsz]\leq n^{-4c}.
$
\end{lemma}
\begin{proof}
We write $|Z_j|=\sum_{i=1}^nY_i$, where $Y_i$ is a random variable indicating whether $i\in Z_j$.
These variables are independent and $\EE[|Z_j|]=\sqrt{n}$. Therefore $\gzsz \ge 2e\sqrt{n} \ge 2e \EE[|Z_j|]$
making Lemma~\ref{lem:chernoff} applicable. This gives
$\Pr[|Z_j|\geq\gzsz]
 \leq2^{-\gzsz}\leq 2^{-4c\log n}=n^{-4c}$.
\end{proof}

We next bound the deviation of $|X\cap Z_j^{(\ell)}|$ from its
expectation $2^\ell k/\sqrt{n}$. At each level, this deviation is
$\Oh(\sqrt{n}\log n)$ with high probability. To state the bound, we define
\begin{equation}\label{eq:gxzsz}
\gxzsz\coloneq16c\sqrt n\log n\leq\Oh(\sqrt n\log n).
\end{equation}

\begin{lemma}\label{lem:rootn-deviation}
    For all $\ell \in [\frac{1}{2}\log n]_0$, $j \in [\sqrt{n}/2^\ell]$,
    \[\Pr\left[\left||X\cap Z^{(\ell)}_j| - \frac{2^\ell k}{\sqrt{n}}\right| \ge \gxzsz\right] \le n^{-4c}.\] 
\end{lemma}
\begin{proof}
If $k<\gxzsz$, both $|X\cap Z_j^{(\ell)}|$ and
$2^\ell k/\sqrt{n}$ lie in $[0,k]$. Their difference in absolute value
is therefore at most $k<\gxzsz$.

Otherwise, $|X\cap Z_j^{(\ell)}|$ is a sum of $k$ independent
Bernoulli variables with mean $\mu=2^\ell k/\sqrt{n}$.
We set $\delta=\sqrt{3(4c+1)\log n/\mu}$. Since
$\mu\geq16c\cdot2^\ell\log n$ and $3(4c+1)<16c$, we have $0<\delta<1$.
Lemma~\ref{lem:chernoff} gives
$\Pr\bigl[\bigl||X\cap Z_j^{(\ell)}|-\mu\bigr|
                \geq\sqrt{3(4c+1)\mu\log n}\bigr]
 \leq2\exp(-(4c+1)\log n)\leq n^{-4c}$.
The last inequality follows from
$2\exp(-(4c+1)\log n)\leq2n^{-(4c+1)}\leq n^{-4c}$,
since logarithms are base two and $n\geq4$.
Finally, $\mu\leq k\leq n$, $3(4c+1)<16c$, and $\log n\geq1$ imply
$\sqrt{3(4c+1)\mu\log n}<4\sqrt{c}\sqrt{n}\log n
<\gxzsz$ by \cref{eq:gxzsz}, proving the claim.
\end{proof}

Note that $|X_B| \le \sqrt{n}$, as every item of $X_B$ has value greater than $t/\sqrt{n}$; if
$|X_B|>\sqrt{n}$, then $\Sigma(X)\geq\Sigma(X_B)>t$, a contradiction.
Removing these items changes the preceding cardinality bound by at most
$\sqrt n$, giving the following bound for $X_A$.
For this, we define
\begin{equation}\label{eq:gazsz}
\gazsz\coloneq20c\sqrt n\log n\leq\Oh(\sqrt n\log n).
\end{equation}

\begin{lemma}\label{lem:XA-size-deviation}
For all $\ell\in[\frac12\log n]_0$ and $j\in[\sqrt n/2^\ell]$,
\[
\Pr\left[\left||X_A\cap Z_j^{(\ell)}|-
\frac{2^\ell k}{\sqrt n}\right|\geq\gazsz\right]\leq n^{-4c}.
\]
\end{lemma}
\begin{proof}
Using $X=X_A\uplus X_B$ and $|X_B| \le \sqrt{n}$, we obtain
\begin{align*}
 \left||X_A\cap Z_j^{(\ell)}|-\frac{2^\ell k}{\sqrt{n}}\right|
 \leq\left||X\cap Z_j^{(\ell)}|-\frac{2^\ell k}{\sqrt{n}}\right|
       +|X_B\cap Z_j^{(\ell)}|
 \leq\left||X\cap Z_j^{(\ell)}|-\frac{2^\ell k}{\sqrt{n}}\right|
       +\sqrt{n}.
\end{align*}
By \cref{eq:gazsz,eq:gxzsz}, since $4c\log n\geq1$, a deviation of at least $\gazsz$
on the left implies a deviation of at least $\gxzsz$
for $X$. Lemma~\ref{lem:rootn-deviation} now gives the desired bound.
\end{proof}

the next lemma states that each color class $Z_j$ contains at most
$\Oh(\log n)$ items from $X_B$, the large items of the fixed solution,
with high probability. 
For this, we first define
\begin{equation}\label{eq:gbzsz}
\gbzsz\coloneq4c\log n\leq\Oh(\log n).
\end{equation}

\begin{restatable}{lemma}{Gbzsz}\label{lem:XB-size-bound}
    For all $j \in [\sqrt{n}]$, we have $\Pr[|X_B \cap Z_j| \ge \gbzsz ] \le n^{-4c}$.
\end{restatable} 

\begin{proof}
    We observe that $\EE[|X_B \cap Z_j|] = |X_B|/ \sqrt{n} \le 1$, where the inequality is because $|X_B| \le \sqrt{n}$. Moreover, $|X_B \cap Z_j| = \sum_{i' \in X_B} Y_{i'}$, where $Y_{i'}$ is the indicator variable of whether $i' \in Z_j$. Since $2e\cdot\EE[|X_B \cap Z_j|] \le \gbzsz$, by Lemma~\ref{lem:chernoff} and \cref{eq:gbzsz} we have $\Pr[|X_B \cap Z_j| \ge \gbzsz] \le 2^{-\gbzsz} = n^{-4c}$.
\end{proof}

For each layer $L_\ell$, we similarly bound the number of solution
items in a color class. The bound is
$\Oh(2^\ell/\sqrt{n}+\log n)$ with high probability. To state it, we define
\begin{equation}\label{eq:gellzsz}
\gellzsz\coloneq\frac{2e\cdot2^\ell}{\sqrt n}+\gbzsz
\leq\Oh(2^\ell/\sqrt n+\log n).
\end{equation}

\begin{lemma}\label{lem:Xj-size-bound}
    $\Pr\left[|X_\ell \cap Z_j| \ge \gellzsz \right] \le n^{-4c}$ for all $j \in [\sqrt{n}]$, $\ell \in \left\{\frac{1}{2}\log n + 1, \ldots, \frac{3}{2}\log n\right\}$.
\end{lemma} 
\begin{proof}
    We observe that $|X_\ell| \le 2^\ell$, as $a_{i'}\in\left(\frac{t}{2^\ell},\frac{t}{2^{\ell-1}}\right]$ for every $i'\in X_\ell$ and $\Sigma(X_\ell) \le \Sigma(X) = t$. Therefore, $\EE[|X_\ell \cap Z_j|] = |X_\ell|/ \sqrt{n} \le 2^\ell/\sqrt{n}$. Moreover, $|X_\ell \cap Z_j| = \sum_{i' \in X_\ell} Y_{i'}$, where $Y_{i'}$ is the indicator variable of whether $i' \in Z_j$. Since $2e\cdot\EE[|X_\ell \cap Z_j|] \le 2^{\ell+1}e/\sqrt{n}\le\gellzsz$ by \cref{eq:gellzsz}, Lemma~\ref{lem:chernoff} gives
    $\Pr\left[|X_\ell \cap Z_j| \ge \gellzsz \right] \le 2^{-\gellzsz}\le 2^{-\gbzsz}=n^{-4c}$.
    The last equality uses \cref{eq:gbzsz}. This proves the lemma.
\end{proof}

We next convert this cardinality bound into a weight bound: for fixed
$\ell$ and $j$, the items in $X_\ell\cap Z_j$ have total weight
$\Oh((t/\sqrt{n})\log n)$ with high probability. 
For this bound, define
\begin{equation}\label{eq:gellzsm}
\gellzsm\coloneq6c\frac{t}{\sqrt n}\log n
\leq\Oh((t/\sqrt n)\log n).
\end{equation}

\begin{restatable}{lemma}{Gellzsm} \label{lem:Xj-sum-bound}
    $\Pr\left[\Sigma(X_\ell \cap Z_j) \ge \gellzsm \right] \le n^{-4c}$ for all $j \in [\sqrt{n}]$, $\ell \in \{\frac{1}{2}\log n + 1, \ldots, \frac{3}{2}\log n\}$.
\end{restatable} 
\begin{proof}
    Note that $2^\ell \ge 2\sqrt{n}$ and $a_{i'}\in\left(\frac{t}{2^\ell},\frac{t}{2^{\ell-1}}\right]$ for every $i'\in X_\ell$. Assume the event $|X_\ell \cap Z_j| < \gellzsz$. This implies $\Sigma(X_\ell \cap Z_j) < \gellzsz\cdot\frac{t}{2^{\ell-1}}$. Substituting the values of $\gellzsz$, $\gbzsz$ and $\gellzsm$ (see \cref{eq:gellzsz,eq:gbzsz,eq:gellzsm}), we get
    \begin{align*}
        \Sigma(X_\ell \cap Z_j) < \gellzsz\cdot\frac{t}{2^{\ell-1}} = \frac{4e\,t}{\sqrt{n}} + \frac{\gbzsz t}{2^{\ell-1}} \le \frac{4e\,t}{\sqrt{n}} + \frac{\gbzsz t}{\sqrt{n}} \le \gellzsm. 
    \end{align*}
    Lemma~\ref{lem:Xj-size-bound} therefore implies
    $\Pr\left[\Sigma(X_\ell \cap Z_j) \ge \gellzsm \right]\le\Pr\left[|X_\ell \cap Z_j| \ge \gellzsz \right] \le n^{-4c}$.
\end{proof}

Summing over the layers and including the smallest items in
$X_C$ gives a bound of $\Oh((t/\sqrt{n})\log^2 n)$ on the total
weight of $X_A\cap Z_j$, with high probability.
We define
\begin{equation}\label{eq:gazsm}
\gazsm\coloneq7c(t/\sqrt n)\log^2 n\leq\Oh((t/\sqrt n)\log^2 n).
\end{equation}

\begin{lemma}\label{lem:XA-sum-bound}
    For $j \in [\sqrt{n}]$, we have $\Pr\left[\Sigma(X_A \cap Z_j) \ge \gazsm \right] \le n^{-3c}$.
\end{lemma} 
\begin{proof}
There are exactly $\log n$ layers $L_\ell$. By
Lemma~\ref{lem:Xj-sum-bound} and a union bound, with probability at
least $1-(\log n)n^{-4c}$, every one of them satisfies $\Sigma(X_\ell\cap Z_j)<\gellzsm$.
Also, deterministically,
$
 \Sigma(X_C\cap Z_j)\leq|X_C|\frac{t}{n\sqrt{n}}
 \leq\frac{t}{\sqrt{n}}
$.
Since $X_A$ is the disjoint union of $X_C$ and the sets $X_\ell$,
these inequalities imply
$
 \Sigma(X_A\cap Z_j)
 < \log n \cdot \gellzsm+\frac{t}{\sqrt{n}}
 \leq\gazsm$.
The last inequality uses \cref{eq:gellzsm,eq:gazsm}.
The probability of this is therefore at least
$1 - (\log n)n^{-4c}\geq 1 - n \cdot n^{-4c}\geq 1 - n^{-3c}$.
\end{proof}

The thresholds used in the preceding bounds are collected in
\cref{tab:color-thresholds} for reference.
\begin{table}[htbp]
\caption{Thresholds in the color-class analysis.}
\label{tab:color-thresholds}
\centering
\small
\begin{tabular}{@{}llll@{}}
\toprule
Threshold & Exact value & Asymptotic bound & Relevant lemma \\
\midrule
$\gzsz$     & $2e\sqrt n+4c\log n$                    & $\Oh(\sqrt n)$ & \cref{lem:Zi-size-bound} \\
$\gxzsz$    & $16c\sqrt n\log n$                    & $\Oh(\sqrt n\log n)$ & \cref{lem:rootn-deviation} \\
$\gazsz$    & $20c\sqrt n\log n$                    & $\Oh(\sqrt n\log n)$ & \cref{lem:XA-size-deviation} \\
$\gbzsz$    & $4c\log n$                            & $\Oh(\log n)$ & \cref{lem:XB-size-bound} \\
$\gellzsz$  & $2e\cdot2^\ell/\sqrt n+4c\log n$    & $\Oh(2^\ell/\sqrt n+\log n)$ & \cref{lem:Xj-size-bound} \\
$\gellzsm$  & $6ct\log n/\sqrt n$                  & $\Oh(t\log n/\sqrt n)$ & \cref{lem:Xj-sum-bound} \\
$\gazsm$    & $7ct\log^2 n/\sqrt n$                & $\Oh(t\log^2 n/\sqrt n)$ & \cref{lem:XA-sum-bound} \\
\bottomrule
\end{tabular}
\end{table}

The next lemma aggregates all proven guarantees into one single statement. It follows by a union bound from
Lemmas~\ref{lem:Zi-size-bound}, \ref{lem:XA-size-deviation}, and~\ref{lem:XA-sum-bound}.

\begin{lemma}[List of probabilistic guarantees]\label{lem:desired-whp-property}
    Over the random choice of the partition of the set $[n]$, 
    all the events below happen
    with probability at least $1 - n^{-3c}$
    \begin{enumerate}
        \item $|Z_j| \le \gzsz\leq\Oh(\sqrt n)$, for $j \in [\sqrt{n}]$,
        \item $\left||X_A\cap Z^{(\ell)}_j| - \frac{2^\ell k}{\sqrt{n}}\right| < \gazsz\leq\Oh(\sqrt n\log n)$, for $\ell \in [\frac{1}{2}\log n]_0$, $j \in [\sqrt{n}/2^\ell]$, and
        \item $\Sigma(X_A \cap Z_j) \le \gazsm\leq\Oh((t/\sqrt n)\log^2 n)$, for $j \in [\sqrt{n}]$.
    \end{enumerate}
\end{lemma}
\begin{proof}
There are $\sqrt{n}$ color classes and $2\sqrt{n}-1$ sets
$Z_j^{(\ell)}$. The first two properties have total failure probability
at most $(3\sqrt{n}-1)n^{-4c}$. The proof of
Lemma~\ref{lem:XA-sum-bound} bounds the failure probability of the
third property, over all color classes, by $\sqrt{n}(\log n)n^{-4c}$.
Thus the total is at most
$(3+\log n)\sqrt{n}\,n^{-4c}\leq n^2n^{-4c}\leq n^{-3c}$,
since $(3+\log n)\sqrt{n}\leq n^2$ for $n\geq4$.
\end{proof}

\subsubsection{Algorithm steps.}

We now have the ingredients to describe our $\OL(n + \sqrt{n}t)$-time 
algorithm for \TCSS. We use \cref{lem:desired-whp-property} to guarantee
certain probabilistic structures. Then, with repeated usage of 
\cref{lem:all-targets-bounded-cardinality}, we compute the
cardinalities and sums of relevant subsets, while pruning out
subsets which are `far from expectation' and hence are ruled out as
irrelevant by \cref{lem:chernoff}.

\begin{theorem}\label{thm:fast-tcss}
For any fixed integer constant $c \geq 10$, there exists a randomized algorithm solving \TCSS on $n$ nonnegative integers with target $t$ in time $\OL(n+\sqrt{n}\,t)$, with error probability at most $n^{-2c}$.
\end{theorem}

\begin{proof}
We first describe the algorithm, prove its correctness, bound its failure
probability, and analyze its running time.

\subparagraph*{Description of the algorithm.}
We form the random color classes and the sets $A,B$ as described in
\cref{sec:col-cod,sec:input-partition}. If some color class $Z_j$ has size more than
$\gzsz$, we return \no.
In every call to \cref{lem:all-targets-bounded-cardinality} below,
we set $\delta=n^{-4c}$ and use fresh randomness.

We apply \cref{lem:all-targets-bounded-cardinality} to $(a_i)_{i\in B}$
with cardinality bound $\sqrt n$ and target $t$, to obtain $\JJ = \{(|S|, \Sigma(S)) \mid S \subseteq B, \Sigma(S) \le t, |S| \le \sqrt{n}\}$. Every subset of $B$ of sum
at most $t$ has fewer than $\sqrt n$ items.

For each $j\in[\sqrt n]$, we apply
\cref{lem:all-targets-bounded-cardinality} to $(a_i)_{i\in Z_j\cap A}$
with cardinality bound $|Z_j\cap A|$ and target
$W=\gazsm$. This gives us
$\II'_j = \{(|S|,\Sigma(S)) : S \subseteq Z_j\cap A, \Sigma(S) \le \gazsm\}$. We define
\[
 \II_j^{(0)}=
 \left\{(d,w):(s,w)\in\II'_j,
 d=s-\left\lfloor\frac{k}{\sqrt{n}}\right\rfloor,
 |d|\leq\gazsz\right\}.
\]
For $1\le\ell\le \frac{1}{2}\log n$, and $1 \le j \le \sqrt{n} / 2^\ell$, we define
\begin{align*}
 \II_j^{(\ell)}=\biggl\{(d,w):\;&
 (d',w)\in\II_{2j-1}^{(\ell-1)}\oplus\II_{2j}^{(\ell-1)},
 d=d'+2\left\lfloor\frac{2^{\ell-1}k}{\sqrt{n}}\right\rfloor
       -\left\lfloor\frac{2^\ell k}{\sqrt{n}}\right\rfloor,
  |d|\leq\gazsz\biggr\}.
\end{align*}
For every $(d, w) \in \II_j^{(\ell)}$, there is a subset of $Z^{(\ell)}_j \cap A$
summing up to $w$ and having size $d + \left\lfloor\frac{2^\ell k}{\sqrt{n}}\right\rfloor$.
 We return \yes\ if some $(s,w)\in\JJ$ satisfies
$(-s,t-w)\in\II_1^{(\frac12\log n)}$, \no\ otherwise.

\subparagraph*{Correctness.}
We first show that the pairs in $\JJ$ and $\II^{(\ell)}_j$
represent cardinalities and sums of existing subsets.
\cref{lem:all-targets-bounded-cardinality} implies that
for every $(s,w)\in\JJ$ there is $S\subseteq B$ with
$|S|=s$ and $\Sigma(S)=w$. Likewise, for every
$(s,w)\in\II'_j$ there is $S\subseteq Z_j\cap A$ with
$|S|=s$ and $\Sigma(S)=w$.

\begin{claim*}
For every $\ell\in[\frac12\log n]_0$, $j\in[\sqrt n/2^\ell]$,
and $(d,w)\in\II_j^{(\ell)}$, there is a subset
$S\subseteq Z_j^{(\ell)}\cap A$ such that
$\Sigma(S)=w$ and
$|S|=d+\lfloor2^\ell k/\sqrt n\rfloor$.
\end{claim*}
\begin{claimproof}
We proceed by induction on $\ell$. If $\ell=0$, the definition of
$\II_j^{(0)}$ gives $(s,w)\in\II'_j$ with
$d=s-\lfloor k/\sqrt n\rfloor$. Therefore,
there is $S\subseteq Z_j\cap A = Z_j^{(0)}\cap A $ with $\Sigma(S)=w$ and
$|S|=s=d+\lfloor k/\sqrt n\rfloor$. 
Now, suppose $\ell\ge1$ and $(d,w)\in\II_j^{(\ell)}$. By construction,
there are $(d_1,w_1)\in\II_{2j-1}^{(\ell-1)}$ and
$(d_2,w_2)\in\II_{2j}^{(\ell-1)}$ with $w=w_1+w_2$, and
 $d=d_1+d_2+2\left\lfloor\frac{2^{\ell-1}k}{\sqrt n}\right\rfloor
          -\left\lfloor\frac{2^\ell k}{\sqrt n}\right\rfloor$.

By the induction hypothesis, there are subsets
$S_1\subseteq Z_{2j-1}^{(\ell-1)}\cap A$ and
$S_2\subseteq Z_{2j}^{(\ell-1)}\cap A$ such that, for
$i\in\{1,2\}$, $\Sigma(S_i)=w_i$ and
$|S_i|=d_i+\lfloor2^{\ell-1}k/\sqrt n\rfloor$.
The sets $Z_{2j-1}^{(\ell-1)}$ and $Z_{2j}^{(\ell-1)}$ are disjoint and
their union is $Z_j^{(\ell)}$. Hence $S_1\cup S_2 \subseteq Z_j^{(\ell)} \cap A$ has sum $w$ and
cardinality $d+\lfloor2^\ell k/\sqrt n\rfloor$, as required.
\end{claimproof}

If the algorithm returns \yes, there are $(s,w)\in\JJ$ and
$(-s,t-w)\in\II_1^{(\frac12\log n)}$. The first pair gives a subset
$S_B\subseteq B$ with $|S_B|=s$ and $\Sigma(S_B)=w$.
Since $Z_1^{(\frac12\log n)}=[n]$, the claim gives a subset
$S_A\subseteq A$ with $|S_A|=-s+k$ and $\Sigma(S_A)=t-w$.
The sets $S_A$ and $S_B$ are disjoint because $A$ and $B$ are disjoint.
Thus $S_A\cup S_B$ has cardinality $k$ and sum $t$, so the algorithm
cannot return \yes\ on a \no-instance.

Now fix a subset $X\subseteq[n]$ with $|X|=k$ and $\Sigma(X)=t$,
independently of the random choices. Let $E$ be the event that all three
properties of \cref{lem:desired-whp-property} hold. Let $E_1$ be the
event that
$\JJ=\{(|S|,\Sigma(S)):S\subseteq B,\ \Sigma(S)\leq t\}$, and let
$E_2$ be the event that
$(|X_A\cap Z_j|,\Sigma(X_A\cap Z_j))\in\II'_j$ for every
$j\in[\sqrt n]$. On $E$, the algorithm does not return \no\ at its
initial check, and $\Sigma(X_A\cap Z_j)\le\gazsm$ for every
$j\in[\sqrt n]$.

We next bound the shifted cardinality of $X_A\cap Z_j^{(\ell)}$
used in the definition of $\II_j^{(\ell)}$. For every
$\ell\in[\frac12\log n]_0$ and $j\in[\sqrt n/2^\ell]$, the event
$E$ implies
\begin{equation}\label{eq:gazsz-bounds-leq}
 \left||X_A\cap Z_j^{(\ell)}|-
       \left\lfloor\frac{2^\ell k}{\sqrt n}\right\rfloor\right|
 \le
 \left||X_A\cap Z_j^{(\ell)}|-\frac{2^\ell k}{\sqrt n}\right|
 +\left(\frac{2^\ell k}{\sqrt n}-
          \left\lfloor\frac{2^\ell k}{\sqrt n}\right\rfloor\right)
 <\gazsz+1.
\end{equation}
Since $c$, $\sqrt n$, and $\log n$ are integers, $\gazsz$ is an
integer. The expression on the left is also an integer, and hence it
is at most $\gazsz$.

\begin{claim*}
If $E$ and $E_2$ both occur, then for every $\ell\in[\frac12\log n]_0$ and
$j\in[\sqrt n/2^\ell]$, we have
\[
 \left(|X_A\cap Z_j^{(\ell)}|-
       \left\lfloor\frac{2^\ell k}{\sqrt n}\right\rfloor,
       \Sigma(X_A\cap Z_j^{(\ell)})\right)\in\II_j^{(\ell)}.
\]
\end{claim*}
\begin{claimproof}
We proceed by induction on $\ell$. If $\ell=0$, the pair
$(|X_A\cap Z_j|,\Sigma(X_A\cap Z_j))$ belongs to $\II'_j$ by $E_2$.
Assuming $E$, \cref{eq:gazsz-bounds-leq} gives
$\bigl||X_A\cap Z_j|-\lfloor k/\sqrt n\rfloor\bigr|\le\gazsz$.
Thus the $\II_j^{(0)}$ contains
$(|X_A\cap Z_j|-\lfloor k/\sqrt n\rfloor,
  \Sigma(X_A\cap Z_j))$.
Now, suppose $\ell\ge1$. By induction hypothesis,
\[
\begin{aligned}
 \left(|X_A\cap Z_{2j-1}^{(\ell-1)}|-
       \left\lfloor\frac{2^{\ell-1}k}{\sqrt n}\right\rfloor,
       \Sigma(X_A\cap Z_{2j-1}^{(\ell-1)})\right)
 &\in\II_{2j-1}^{(\ell-1)},\\
 \left(|X_A\cap Z_{2j}^{(\ell-1)}|-
       \left\lfloor\frac{2^{\ell-1}k}{\sqrt n}\right\rfloor,
       \Sigma(X_A\cap Z_{2j}^{(\ell-1)})\right)
 &\in\II_{2j}^{(\ell-1)}.
\end{aligned}
\]
The componentwise sum of these two pairs is $(d',w)$, where
\[
 d'=|X_A\cap Z_j^{(\ell)}|-
     2\left\lfloor\frac{2^{\ell-1}k}{\sqrt n}\right\rfloor,
 \qquad w=\Sigma(X_A\cap Z_j^{(\ell)}).
\]
Substituting this value of $d'$ into the defining relation for
$\II_j^{(\ell)}$ gives
\begin{equation}\label{eq:XA-pair-at-level}
 \left(|X_A\cap Z_j^{(\ell)}|-
       \left\lfloor\frac{2^\ell k}{\sqrt n}\right\rfloor,
       \Sigma(X_A\cap Z_j^{(\ell)})\right).
\end{equation}
Again, \cref{eq:gazsz-bounds-leq} shows that
$\bigl||X_A\cap Z_j^{(\ell)}|-
\lfloor2^\ell k/\sqrt n\rfloor\bigr|\le\gazsz$, implying
that the pair in \cref{eq:XA-pair-at-level} belongs to $\II_j^{(\ell)}$.
\end{claimproof}

At $\ell=\frac12\log n$ and $j=1$, we have
$Z_1^{(\ell)}=[n]$ and
$\lfloor2^\ell k/\sqrt n\rfloor=k$. Thus, on $E\cap E_2$,
$(|X_A|-k,\Sigma(X_A))\in\II_1^{(\frac12\log n)}$.
On $E_1$, the pair $(|X_B|,\Sigma(X_B))$ belongs to $\JJ$.
Because $|X_A|-k=-|X_B|$ and
$\Sigma(X_A)=t-\Sigma(X_B)$, these two pairs satisfy the
acceptance condition, and the algorithm returns \yes.

\subparagraph*{Failure probability.}
By \cref{lem:desired-whp-property}, $\Pr[\neg E]\leq n^{-3c}$.
By \cref{lem:all-targets-bounded-cardinality}, each call fails with
probability at most $\delta=n^{-4c}$. Thus
$\Pr[\neg E_1]\leq n^{-4c}$. For every partition satisfying $E$, each of the
$\sqrt n$ calls for $\II'_j$ contains its eligible pair with failure
probability at most $n^{-4c}$. A union bound gives
$\Pr[\neg E_2\mid E]\leq\sqrt n\,n^{-4c}\leq n^{-3c}$.
Hence the total failure probability is at most
$n^{-3c}+n^{-4c}+n^{-3c}
 \leq3n^{-3c}\leq n^{-2c}$.

\subparagraph*{Running time.}
The initial partition into color classes and preprocessing take $\Oh(n)$ time.
Since $\log(1/\delta)=4c\log n$, the dependence on $\delta$ in
\cref{lem:all-targets-bounded-cardinality} contributes only
polylogarithmic factors.
Our initial check based on sizes of color classes ensures that every color class has at most $\gzsz$
items whenever we perform the remaining computations. The call to \cref{lem:all-targets-bounded-cardinality} for computing
$\JJ$ takes $\OL(|B|+\sqrt n\,t)$ time. By \cref{eq:gazsm}, the call
to \cref{lem:all-targets-bounded-cardinality} for computing $\II'_j$ takes $\OL(|Z_j|+|Z_j|t/\sqrt n)$ time. In total, computing $\II^{(0)}_j$ for all $j \in [\sqrt{n}]$ takes
\[
 \sum_{j=1}^{\sqrt n}\OL(|Z_j|+|Z_j|t/\sqrt n)
 =\OL(n+\sqrt n\,t)
\]
time. Next we compute $\II^{(\ell)}_j$ for $\ell \ge 1$ using \cref{lem:pair-sumset}. We bound $w$ for all $(d,w) \in \II^{(\ell)}_j$.

\begin{claim}
For $\ell\in[\frac12\log n]_0$, $j\in[\sqrt n/2^\ell]$, and
$(d,w)\in\II_j^{(\ell)}$, we have $0 \le w \le 2^\ell\gazsm$.
\end{claim}
\begin{claimproof}
We proceed by induction on $\ell$. If $\ell=0$, the definition of
$\II_j^{(0)}$ gives $(s,w)\in\II'_j$ for some $s$. Every pair in
$\II'_j$ is realized by a subset of $Z_j\cap A$ with nonnegative sum
at most $\gazsm$, so $0\le w\le\gazsm$. Now, suppose $\ell\ge1$. By the definition of $\II_j^{(\ell)}$, some
$(d',w)$ belongs to
$\II_{2j-1}^{(\ell-1)}\oplus\II_{2j}^{(\ell-1)}$.
Thus there are $(d_1,w_1)\in\II_{2j-1}^{(\ell-1)}$ and
$(d_2,w_2)\in\II_{2j}^{(\ell-1)}$ with $w=w_1+w_2$.
By the induction hypothesis, $0\le w_1,w_2\le2^{\ell-1}\gazsm$.
Hence $0\le w\le2^\ell\gazsm$.
\end{claimproof}

Fix $\ell\in\{1,\ldots,\frac12\log n\}$ and
$j\in[\sqrt n/2^\ell]$. Every $(d,w)$ in either
$\II_{2j-1}^{(\ell-1)}$ or $\II_{2j}^{(\ell-1)}$ satisfies
$|d|\le\gazsz$ by definition and
$0\le w\le2^{\ell-1}\gazsm$ by the claim. Therefore,
\cref{lem:pair-sumset} computes their sumset in time
$\OL\bigl(\gazsz(2^{\ell-1}\gazsm+1)\bigr)
 =\OL(2^\ell t+\sqrt n)$
by \cref{eq:gazsz,eq:gazsm}. Every pair $(d',w)$ in this sumset has
$|d'|\le2\gazsz$ and $0\le w\le2^\ell\gazsm$, so the sumset contains
$\Oh(\gazsz(2^\ell\gazsm+1))$ pairs. Scanning them to form
$\II_j^{(\ell)}$ also takes $\OL(2^\ell t+\sqrt n)$ time.
There are $\sqrt n/2^\ell$ such computations for this value of
$\ell$, so their total time over all $\ell$ is
\[
 \sum_{\ell=1}^{\frac12\log n}\frac{\sqrt n}{2^\ell}
        \OL(2^\ell t+\sqrt n)=\OL(n+\sqrt n\,t).
\]
Finally, we store $\II_1^{(\frac12\log n)}$ in a Boolean array
indexed by $(d,w)$. Its definition gives $|d|\le\gazsz$, and the
claim gives $0\le w\le\sqrt n\,\gazsm$; thus initializing and populating the array
takes $\OL(\gazsz(\sqrt n\,\gazsm+1))=\OL(\sqrt n\,t)$ time.
The set $\JJ$ has at most $(\sqrt n+1)(t+1)$ pairs, so scanning it
and checking the acceptance condition takes $\OL(\sqrt n\,t)$ time.
This proves the theorem.
Algorithm~\ref{alg:fasttcss} summarizes the construction.
\end{proof}

\begin{algorithm}[!htbp]
\caption{$\fasttcss((a_1,\ldots,a_n),t,k)$; see \cref{thm:fast-tcss}}\label{alg:fasttcss}
\begin{algorithmic}[1]
 \State Replace values greater than $t$ by $t+1$ and add values of $t+1$ to get a power-of-four size.
 \State Define $A,B$ as above. \Comment{see \cref{sec:input-partition}}
 \State $\gzsz\gets2e\sqrt n+4c\log n$, $\gazsz\gets20c\sqrt n\log n,\gazsm\gets7c(t/\sqrt n)\log^2 n, \delta\gets n^{-4c}$
 \State $Z_1,\ldots,Z_{\sqrt{n}}\gets\varnothing$ \Comment{Color classes; see \cref{sec:col-cod}}
 \For{$i=1,\ldots,n$}
   \State Choose $j$ independently and uniformly from $[\sqrt{n}]$.
   \State $Z_j\gets Z_j\cup\{i\}$
 \EndFor
  \If{some $|Z_j|>\gzsz$}
   \State \Return \no
 \EndIf
 \For{$j=1,\ldots,\sqrt{n}$}
   \State $\II'_j\gets\{(|S|,\Sigma(S)):S\subseteq Z_j\cap A,\ \Sigma(S)\leq\gazsm\}$
          \Comment{Lemma~\ref{lem:all-targets-bounded-cardinality}}
   \State $\II_j^{(0)}\gets\{(s-\lfloor k/\sqrt{n}\rfloor,w):(s,w)\in\II'_j,
                \ |s-\lfloor k/\sqrt{n}\rfloor|\leq\gazsz\}$
 \EndFor
 \State $\JJ\gets\{(|S|,\Sigma(S)):S\subseteq B,\ \Sigma(S)\leq t\}$
        \Comment{Lemma~\ref{lem:all-targets-bounded-cardinality}}
 \For{$\ell=1,\ldots,\frac12\log n$}
   \For{$j=1,\ldots,\sqrt{n}/2^\ell$}
     \State $\II'\gets\II_{2j-1}^{(\ell-1)}\oplus\II_{2j}^{(\ell-1)}$
     \State $\II_j^{(\ell)}\gets\left\{(d,w):(d',w)\in\II',
                       |d|\leq\gazsz, d=d'+2\left\lfloor\frac{2^{\ell-1}k}{\sqrt{n}}\right\rfloor
                   -\left\lfloor\frac{2^\ell k}{\sqrt{n}}\right\rfloor\right\}$
   \EndFor
 \EndFor
 \For{$(s,w)\in\JJ$}
   \If{$(-s,t-w)\in\II_1^{(\frac12\log n)}$}
     \State \Return \yes
   \EndIf
 \EndFor
 \State \Return \no
\end{algorithmic}
\end{algorithm}

\subsection{Algorithm for Largest Cardinality Subset Sum}

\TCSS asks to decide if there exists a subset with sum $t$ and cardinality
$k$. Our next step to solve \LCSS is to decide if there exists a subset with
sum $t$ and cardinality in an interval $[\ell, h]$. We show that this can be
done by simply adding $h - \ell$ zeros to the instance and asking whether a 
subset exists with sum $t$ and cardinality $h$.

\begin{lemma}[Target cardinality to cardinality in an interval]\label{lem:add-zeros}
Let $a_1,\ldots,a_n$ and $t$ be nonnegative integers, and let
$0\leq\ell\leq h$ be integers. Then the following are equivalent:
\begin{enumerate}
 \item There is a subset $S\subseteq[n]$ with $\Sigma(S)=t$ and
       $|S|\in[\ell,h]$.
 \item The \TCSS instance obtained by appending $h-\ell$ zero-valued items,
       with target cardinality $h$ and target $t$, is a \yes-instance.
\end{enumerate}
\end{lemma}
\begin{proof}
Suppose the first statement holds, and let $S\subseteq[n]$ have
$\Sigma(S)=t$ and $|S|\in[\ell,h]$. Since $h-|S|\leq h-\ell$, we can
add $h-|S|$ of the appended zeros to $S$. The resulting subset has
cardinality $h$ and sum $t$, proving the second statement.
Conversely, suppose the second statement holds, and let
$S'\subseteq[n+h-\ell]$ have cardinality $h$ and sum $t$ in the
augmented instance. The subset $S=S'\cap[n]$ has sum $t$, since all
appended items have value zero. Moreover,
 $|S|=h-|S'\cap\{n+1,\ldots,n+h-\ell\}|
     \geq h-(h-\ell)=\ell$,
while $|S|\leq|S'|=h$. Thus $S$ satisfies the first statement.
\end{proof}

This allows us to solve \LCSS using a binary-search approach.

\fasLSSalgo*

\begin{proof}
First we test feasibility by calling \cref{thm:fast-tcss} with target $t$ and
target cardinality $n$ on the original items augmented by $n$ zeros.
A correct call returns
\yes\ exactly when the original instance has a subset of sum $t$ and
cardinality in $[1,n]$. Since $t>0$, this is equivalent to the
existence of a subset of sum $t$. If the call returns \no, we output
that no solutions exist. The test takes
$\OL(n+\sqrt n\,t)$ time and never declares an infeasible instance
feasible.

If the feasibility test succeeds, set
$\ell=0$, $h=n+1$. Let $k^*$ be the cardinality of the largest subset summing up to $t$. Our invariant throughout will be $k^* \in [\ell, h)$; this holds at initialization.
At each iteration, we set $m=\lfloor(\ell+h)/2\rfloor$ and apply
\cref{thm:fast-tcss} with target cardinality $h$ to the original items
augmented by $h-m$ zeros. By \cref{lem:add-zeros}, a correct
\yes\ answer occurs exactly when the original instance has a subset
of sum $t$ and cardinality in $[m,h]$. From the invariant in the previous iteration, we had $k^* < h$. Thus a correct \yes\ answer implies
$k^*\in[m,h)$, and a correct \no\ answer implies $k^*\in[\ell,m)$.
We update $\ell \gets m$ and $h \gets m$ respectively for these cases, preserving the invariant. When $h=\ell+1$, the invariant gives $k^*=\ell$.
Including the feasibility test, there are at most
$(1+\lceil\log(n+1)\rceil)$ calls to \cref{thm:fast-tcss}, each on at most $2n$
items. Their total time is
$\OL(n+\sqrt{n}\,t)$. 
Conditioning on the history before any call fixes its input, 
\cref{thm:fast-tcss} bounds its conditional error probability of any specific call by
$n^{-2c}$. A union bound gives
total error at most $(1+\lceil\log(n+1)\rceil)n^{-2c}
\leq n^{-c}$.
Algorithm~\ref{alg:LSS-algo} summarizes this.
\end{proof}

\begin{algorithm}[!htbp]
\caption{$\mathtt{FasterLCSS}((a_1,\ldots,a_n),t)$; see \cref{thm:fas-LSS-algo}}\label{alg:LSS-algo}
\begin{algorithmic}[1]
 \If{$\fasttcss((a_1,\ldots,a_n,\underbrace{0,\ldots,0}_{n}),t,n)=\no$} \Comment{\cref{alg:fasttcss}}
   \State \Return ``$\texttt{No Subset with sum }t$''
 \EndIf
 \State $\ell\gets0$, $h\gets n+1$
 \While{$\ell<h-1$}
   \State $m\gets\lfloor(\ell+h)/2\rfloor$
   \If{$\fasttcss((a_1,\ldots,a_n,\underbrace{0,\ldots,0}_{h-m}),t,h)=\yes$}
     \State $\ell\gets m$
   \Else
     \State $h\gets m$
   \EndIf
 \EndWhile
 \State \Return $\ell$
\end{algorithmic}
\end{algorithm}

\section{All-Targets Largest Cardinality Subset Sum}
\label{sec:all-targets}

We recall the problem of \ATLCSS (\cref{def:prob-AT-LCSS}):
Given nonnegative integers $a_1,\ldots,a_n$ and a positive integer $t$, the task is to
compute, for \emph{every} $s\in[t]$, the largest cardinality of
a subset of indices whose sum is $s$. 
For this section, we assume without loss of generality that all items are positive 
(otherwise we simply delete all zero items and add back their count in all outputs, 
see proof of \cref{thm:all-LSS-algo}).

We fix an integer threshold $1\leq\tau\leq t$. We define heavy and light elements as follows.

\begin{definition}[Heavy elements $H$ and light elements $L$]\label{def:HL}
    Let $H = \{i \in [n] : a_i > \tau\}$ be heavy elements, and let $L = \{i \in [n] : a_i \le \tau\}$ 
    be light elements.
\end{definition}

We handle heavy items and light items separately, and then incorporate them to get an 
efficient algorithm for \ATLCSS.

\subsection{Handling heavy items}

A simple application of \cref{lem:all-targets-bounded-cardinality} 
allows us to efficiently compute and store for every $s \in [t]_0$
the largest cardinality of a subset of the heavy element indices which sum up to $s$.

\begin{lemma}\label{lem:all-targets-heavy}
There exists a randomized algorithm computing $\mathtt{largest}(s)=\max\{|S|:S\subseteq H,\ \Sigma(S)=s\}$ for every $s\in[t]_0$
in time $\OL(|H|+t^2/\tau)$ with failure probability $\le n^{-c}$.
\end{lemma}
\begin{proof}
Each item in $H$ is greater than $\tau$. Hence every subset
with sum at most $t$ has cardinality less than $t / \tau$.
We apply Lemma~\ref{lem:all-targets-bounded-cardinality} once to
$(a_i)_{i\in H}$ with $k=\lfloor t/\tau\rfloor$ and
$\delta=n^{-c}$ to obtain
$\mathcal A = \{(\Sigma(S), |S|) \mid S \subseteq H, |S| \le k, \Sigma(S) \le t\}$
with probability at least $1-n^{-c}$.
Every returned pair is realized by a subset of $H$. Since
$\log(1/\delta)=c\log n$, the call takes
$\OL(|H|+kt)=\OL(|H|+t^2/\tau)$ time.

We initialize the array $\mathtt{largest}(\cdot)$ to $-\infty$ and, for
each returned pair $(s,r) \in \mathcal A$, we replace $\mathtt{largest}(s)$ by the maximum
of its current value and $r$; the resulting entry is precisely the
largest cardinality of a subset $S \subseteq H$ with sum $s$.
\end{proof}

For each sum $s \in [t]_0$, it is sufficient to store only this maximum
cardinality $\mathtt{largest}(s)$. All items subsequently added come from the light
items in the input.

\subsection{Processing light items}

We consider all items in the input to have an associated \emph{profit} of $1$.
Then \ATLCSS asks for every $s \in [t]$,
the maximum profit subset with values summing up to $s$.

We now represent items by a pair $(w,p)$, where $w$ is its value and $p$
is its profit. We consider the light items in the input. Initially, each light item $a_i$ is represented by
$(a_i,1)$. We now use the multiplicity reduction trick from
Lawler~\cite{lawler1979fast}, which deletes some of the existing items and 
introduces new items (possibly with profit more than $1$) such that the maximum profit of subsets summing up to some $s \in [t]$
does not change. This leads to a reduced total number of light items. We formally state it in the following lemma.

\begin{lemma}[Small number of (weight, profit) pairs from light items]\label{lem:all-targets-light}
In time $\Oh(|L| + \tau \log t)$, one can construct a
multiset $\mathcal L$ of size at most $\Oh(\tau \log t)$, which
contains pairs of positive integers,
such that for every $s \in [t]_0$
\begin{equation}\label{eq:wp-pairs}
 \max\{|S|: S \subseteq L, \Sigma(S) = s\}
 = \max\!\left\{\sum_{(w,p)\in S'}p : S'\subseteq\mathcal L,
      \sum_{(w,p)\in S'}w=s\right\}.
\end{equation}
\end{lemma}
\begin{proof}
For each pair $(w,p)\in \mathcal L$, we refer to $w$ as its \emph{weight} and to $p$ as its \emph{profit}.
We first describe the construction of $\mathcal L$, prove that \cref{eq:wp-pairs} holds, then bound the size of $\mathcal L$ and finally argue about the time complexity to compute $\mathcal L$.

\subparagraph*{Construction of $\mathcal L$.}
Initially, for each index $i\in L$ we add to $\mathcal L$ the pair $(a_i,1)$.
We repeatedly perform the following operations. We stop when neither applies.
\begin{enumerate}
 \item If a pair $(w,p)$ occurs at least three times, we remove two copies and insert one copy of $(2w,2p)$.
 \item We delete every pair $(w, p)$ with $w > t$.
\end{enumerate}

\subparagraph*{Proof of \cref{eq:wp-pairs}.}
We prove that for every $s\in[t]_0$, the value of
\[
\max\!\left\{\sum_{(w,p)\in S'}p : S'\subseteq\mathcal L,
  \sum_{(w,p)\in S'}w=s\right\}
\]
remains unchanged after each iteration, where $\mathcal L$ denotes the current
multiset. Fix $s \in [t]_0$.
We suppose that in some iteration we replace two copies of a pair
$(w,p)$ occurring at least three times by one copy of $(2w,2p)$.
We compare the value of this maximum on the multisets
$\mathcal L_{\mathrm{before}}$ and $\mathcal L_{\mathrm{after}}$ immediately before
and after this operation, respectively. We refer to all copies of $(w, p)$ which were not removed in this iteration as \emph{surviving copies}.

The maximum cannot increase: given any submultiset of
$\mathcal L_{\mathrm{after}}$ containing the newly inserted copy of $(2w,2p)$,
we replace that copy by the two copies of $(w,p)$ removed from
$\mathcal L_{\mathrm{before}}$. These two copies are distinct from all
surviving copies, so this replacement respects multiplicities and
preserves both total weight and total profit. On the other hand, a submultiset that does
not contain the newly inserted copy is already available in
$\mathcal L_{\mathrm{before}}$.

The maximum cannot decrease either. Given any submultiset of
$\mathcal L_{\mathrm{before}}$ containing at least two copies of $(w,p)$,
we replace two of its copies by the newly inserted copy of $(2w,2p)$.
We can select its remaining copies of $(w,p)$ from those surviving
in $\mathcal L_{\mathrm{after}}$. If the submultiset
contains exactly one copy of $(w,p)$, we select a surviving copy;
such a copy exists because there were at least three copies before
the operation. If it contains no copies of $(w,p)$, we leave it
unchanged. In each case, we retain all other selected pairs and
preserve total weight and total profit.

Each iteration decreases the number of pairs by at least one. Hence the construction
terminates after at most $|L|$ such iterations.
Initially, we have one pair $(a_i,1)$ for each index $i\in L$.
Thus subsets of $L$ correspond to submultisets of the initial pairs
with the same total weight and with total profit equal to their
cardinality, establishing \cref{eq:wp-pairs} at initialization.
Since each iteration preserves the maximum in \cref{eq:wp-pairs}, the equality
holds for the final multiset $\mathcal L$ as well.

\subparagraph*{Size of $\mathcal L$.}
We first observe that every pair occurring during the construction has
the form $(2^jv,2^j)$ for some $v\in[\tau]$, $j\geq0$. This is because
initially this holds with $j=0$ and subsequently if we replace two copies of
$(2^jv,2^j)$, the inserted pair is $(2^{j+1}v,2^{j+1})$.

At termination, every pair in $\mathcal L$ has weight at most $t$ and multiplicity at most two.
For a pair $(2^jv,2^j)$ in $\mathcal L$, its weight is at most $t$, so
$0\leq j\leq\lfloor\log(t/v)\rfloor$. For each $v\in[\tau]$,
there are thus $\lfloor\log(t/v)\rfloor+1$ possible pairs,
each of multiplicity at most two. This gives us $|\mathcal L|=\Oh(\tau\log t)$.

\subparagraph*{Running time.}
For completeness, we describe an efficient implementation of the
construction of $\mathcal L$. We start by
storing the multiplicity of each possible pair $(2^jv,2^j)$ in a
two-dimensional array indexed by $v\in[\tau]$ and
$0\leq j\leq\lfloor\log(t/v)\rfloor$. Distinct indices represent
distinct pairs: the profit determines $j$, and the weight then
determines $v$. We initialize the array to zero and count the input
occurrences in the entries with $j=0$, in time
$\Oh(|L|+\tau\log t)$.

For each $v$, we scan its entries in increasing order of $j$.
While the current multiplicity is at least three, we (i) decrease it by
two and, (ii) if $2^{j+1}v\leq t$, increase the next entry (i.e.~indexed 
by $v$ and $j + 1$) by one. Each such step decreases
the total number of items by at least one, so there are at most $|L|$
steps in total. Since updates only affect the entries with a larger 
value of $j$, each entry which has already been processed has a multiplicity of at
most two. We output these remaining copies. Scanning the array and
writing the output take $\Oh(\tau\log t)$ time. Hence the total running
time is bounded by $\Oh(|L|+\tau\log t)$.
\end{proof}

\subsection{Algorithm for All-Targets Largest Cardinality Subset Sum}

We now have all ingredients to prove \cref{thm:all-LSS-algo}.

\allLSSalgo*

\begin{proof}

Let $z=|\{i\in[n]:a_i=0\}|$ be the number of items with value $0$. 
We now discard all items of value $0$ (we simply add $z$ to every 
output in the end) and all items of value greater than $t$. We set 
$\tau=\lceil\sqrt{t}\rceil$. Let $L$ and $H$ be the sets of indices
of remaining items with values at most $\tau$ and greater than
$\tau$, respectively (recall \cref{def:HL}). 

By \cref{lem:all-targets-heavy}, we compute
$\mathtt{largest}(s)$ for every $s\in[t]_0$, where
$\mathtt{largest}(s)$ is the maximum cardinality of a subset of
$H$ whose sum is $s$. By \cref{lem:all-targets-light}, we construct a
multiset $\mathcal L$ of weight--profit pairs such that, for every
$s\in[t]$, the maximum profit of a submultiset of $\mathcal L$ of
weight $s$ equals the maximum cardinality of a subset of $L$ of
sum $s$. We enumerate the occurrences in $\mathcal L$ as
$(w_1,p_1),\ldots,(w_m,p_m)$, where $m=|\mathcal L| = \Oh(\tau\log t)$.
For the correctness proof, we condition on the event that the call to \cref{lem:all-targets-heavy} was successful. This happens with probability at least $1 - n^{-c}$.

We approach using dynamic programming. The entry $\DP_j[s]$ stores the maximum of the number of selected heavy
items plus the profit of selected pairs among the first $j$ pairs
of $\mathcal L$, subject to their total weight being $s$. Formally, for
$0\leq j\leq m$ and $0\leq s\leq t$, we define
\[
 \DP_j[s]:=
 \max\left\{|S|+\sum_{i\in I}p_i :
 S\subseteq H,\ I\subseteq[j],
 \Sigma(S)+\sum_{i\in I}w_i=s\right\}.
\]
We initialize
$\DP_0[s]=\mathtt{largest}(s)$ and, for $j=1,\ldots,m$, we compute
\[
 \DP_j[s]=
 \begin{cases}
  \DP_{j-1}[s],&s<w_j,\\
  \max\{\DP_{j-1}[s],\ \DP_{j-1}[s-w_j]+p_j\},&s\geq w_j
 \end{cases}.
\]

We prove by induction on $j$ that the recurrence computes the stated
value of $\DP_j[s]$ for every $s\in[t]_0$. For $j=0$, no pair
from $\mathcal L$ is available, so $\DP_0[s]=\mathtt{largest}(s)$ by their definitions. Now let
$j\geq1$ and assume the claim for $j-1$. If $s<w_j$, then a subset of
the first $j$ pair indices having total weight at most $s$ cannot contain the
$j$-th pair $(w_j,p_j)$.
Consequently, $\DP_j[s]=\DP_{j-1}[s]$. From now, we suppose that $s\geq w_j$. 

For the upper bound, we consider
$S^*\subseteq H$ and $I^*\subseteq[j]$ satisfying
$\Sigma(S^*)+\sum_{i\in I^*}w_i=s$ such that $\DP_j[s] = |S^*| + \sum_{i \in I^*} p_i$. 
If $j\notin I^*$, then
$|S^*|+\sum_{i\in I^*}p_i\leq\DP_{j-1}[s]$. If $j\in I^*$, then
$\Sigma(S^*)+\sum_{i\in I^*\setminus\{j\}}w_i=s-w_j$, and hence
$|S^*|+\sum_{i\in I^*}p_i\leq\DP_{j-1}[s-w_j]+p_j$. As at least one of these
inequalities holds, we get 
\begin{equation}\label{eq:dp-upper-bound}
    \DP_j[s] = |S^*|+\sum_{i\in I^*}p_i\leq\max\{\DP_{j-1}[s],\DP_{j-1}[s-w_j]+p_j\}.
\end{equation}

For the reverse inequality, we consider
$S_1^*\subseteq H$ and $I_1^*\subseteq[j]\setminus\{j\}$ satisfying
$\Sigma(S_1^*)+\sum_{i\in I_1^*}w_i=s$ such that
$\DP_{j-1}[s]=|S_1^*|+\sum_{i\in I_1^*}p_i$.
Since these subsets are also considered in the definition of $\DP_j[s]$,
we get $\DP_j[s]\geq |S_1^*|+\sum_{i\in I_1^*}p_i=\DP_{j-1}[s]$.
Similarly, we consider $S_2^*\subseteq H$ and
$I_2^*\subseteq[j]\setminus\{j\}$ satisfying
$\Sigma(S_2^*)+\sum_{i\in I_2^*}w_i=s-w_j$ such that
$\DP_{j-1}[s-w_j]=|S_2^*|+\sum_{i\in I_2^*}p_i$.
We retain $S_2^*$ and replace $I_2^*$ by $I_2^*\cup\{j\}$, thereby
selecting the $j$-th pair $(w_j,p_j)$. The total weight becomes $s$, so
$\DP_j[s]\geq |S_2^*|+\sum_{i\in I_2^*}p_i+p_j=\DP_{j-1}[s-w_j]+p_j$.
Together, these inequalities give
\begin{equation}\label{eq:dp-lower-bound}
 \DP_j[s]\geq
 \max\{\DP_{j-1}[s],\DP_{j-1}[s-w_j]+p_j\}.
\end{equation}
The two inequalities \cref{eq:dp-upper-bound,eq:dp-lower-bound} give
the desired equality and complete the induction.

It remains to relate $\DP_m$ to the original instance. For every
$s\in[t]$, the definition of $\DP_m[s]$ gives
\begin{align*}
 \DP_m[s] &= \max\left\{|S|+\sum_{(w,p)\in S'}p :
 S\subseteq H,\ S'\subseteq\mathcal L,
 \Sigma(S)+\sum_{(w,p)\in S'}w=s\right\} \\
 &=\max_{r \in [s]_0}\left\{
     \mathtt{largest}(r)+
     \max\left\{\sum_{(w,p)\in S'}p:
       S'\subseteq\mathcal L,\ \sum_{(w,p)\in S'}w=s-r\right\}
   \right\}\\
 &=\max_{r \in [s]_0}\big\{
     \max\{|S_1|:S_1\subseteq H,\ \Sigma(S_1)=r\}
     +\max\{|S_2|:S_2\subseteq L,\ \Sigma(S_2)=s-r\}
   \big\}\\
 &=\max\{|S|:S\subseteq H\cup L,\ \Sigma(S)=s\}.
\end{align*}
The first equality is the definition of $\DP_m[s]$. For the second
equality, we group the subsets by $r=\Sigma(S)$ and maximize the heavy
cardinality and light profit independently for each $r$.
The third equality follows from the definition of
$\mathtt{largest}(r)$ and \cref{eq:wp-pairs}. The fourth equality
follows from observing that subsets $S_1 \subseteq H$ and $S_2 \subseteq L$
form precisely the subsets of $H\cup L$ which sum up to $s$.
Every subset of $H\cup L$ of sum $s$ can be enlarged by all $z$
zero-valued items. Conversely, deleting zero-valued items from any
original solution of sum $s$ leaves a subset of $H\cup L$ of the same
sum. Thus the required answer is
$\DP_m[s]+z$ if $\DP_m[s]$.

Preprocessing takes $\Oh(n)$ time. The two lemmas take
$\OL(|H|+t^2/\tau)$ and $\Oh(|L|+\tau\log t)$ time,
respectively. The dynamic program takes $\Oh(mt) = \Oh(\tau t \log t)$ time, and
output takes $\Oh(t)$ time. Since $|L|+|H|\leq n$
and $\tau = \Oh(\sqrt{t})$, the total is
\[
 \OL\left(n+\frac{t^2}{\tau}+\tau t\right)
 =\OL(n+t^{3/2}).
\]
The only randomized step is the heavy-item computation, which is
correct for all targets simultaneously with probability at least
$1-n^{-c}$ by \cref{lem:all-targets-heavy}.
\end{proof}
 \section{Projection Subset Sum}\label{sec:PSS}

We consider the search version of Projection Subset Sum, introduced by
Dutta and Rajasree~\cite{DR23}.

Throughout this section, $Z$ and $X$ denote the vectors
$(z_1,\ldots,z_n)$ and $(x_1,\ldots,x_n)$, respectively.
For an input $a_1,\ldots,a_n$, we set $z_i=a_i$ for every $i\in[n]$.

\begin{definition}[\textsf{Projection Subset Sum} $(\PSS_p)$]
\label{def:pss-search}
Fix $p\geq1$. Given nonnegative integers
$Z=(a_1,\ldots,a_n)$ and a nonnegative integer $t$, the $\PSS_p$
problem asks for a vector $X=(x_1,\ldots,x_n)\in\Z^n$ and a set
$S\subseteq[n]$ such that $\sum_{i\in S}x_i=t$ and
$\norm{Z-X}_p$ is minimized.
\end{definition}

We assume $n\geq1$ and write
$\Sigma(S)=\sum_{i\in S}a_i$.
A sum $s$ is attainable if
$\Sigma(S)=s$ for some $S\subseteq[n]$. When $t=0$, we return $X=Z$ and
$S=\emptyset$. When $n=1$ and $t>0$, we return $X=(t)$ and $S=\{1\}$.
The proofs of the two $\PSS$ theorems below therefore assume $n\geq2$
and $t>0$, so every feasible set is nonempty.

\subsection{Projection Subset Sum for all norms}

Before presenting our algorithm for $\PSS_p$, we first show how to use the $\ATLCSS$ algorithm of \cref{thm:all-LSS-algo} to find a largest subset whose elements sum to a prescribed target.
When applying this theorem to items indexed by $I\subseteq[n]$,
we denote its output by $L$ and include
$L(0)=|\{i\in I:a_i=0\}|$.

\begin{lemma}\label{lem:pss-largest-subset}
Given nonnegative integers $a_1,\ldots,a_n$ and an integer $t\geq0$,
there is a randomized algorithm that, with probability at least $5/6$,
outputs a largest subset $S\subseteq[n]$ with $\Sigma(S)=t$, or
$\fail$ if no such subset exists, in time $\OL(n+t^{3/2})$.
\end{lemma}
\begin{proof}
We first describe the algorithm, prove its correctness, bound its failure
probability, and analyze its running time.

\subparagraph*{Description of the algorithm.}
Start with $I=[n]$ and $u=t$. On a recursive instance $(I,u)$,
return $\{i\in I:a_i=0\}$ if $u=0$.
For $I=\{i\}$ and $u>0$, return $\{i\}$ if
$a_i=u$, and $\fail$ otherwise.

For $|I|\geq2$ and $u>0$, split $I=I_1\uplus I_2$ with
$\bigl||I_1|-|I_2|\bigr|\leq1$. For each $r\in\{1,2\}$, append
one item of value $u+1$ to $(a_i)_{i\in I_r}$ and solve \ATLCSS
with sum cap $u$ using \cref{thm:all-LSS-algo} with $c=10$ (the addition of the item simply to ensures that the number of items in any recursive call is at least $2$, hence the failure probability is at most $2^{-10}$).
The added item cannot occur in a subset of sum at most $u$.
Execute this algorithm independently
$R=2\lceil\log_2(12n)\rceil+1$ times and let $L_r(s)$ be the median
of its $R$ outputs (across the $R$ independent runs) for each $s\in[u]_0$.
Scan $s\in[u]_0$ with $L_1(s)\geq0$ and $L_2(u-s)\geq0$,
and choose one maximizing $L_1(s)+L_2(u-s)$.
If no such $s$ exists, return $\fail$.
Apply the same procedure to $(I_1,s)$ and $(I_2,u-s)$.
If either invocation returns $\fail$, return $\fail$; otherwise,
return the union of the two subsets it outputs.

\subparagraph*{Correctness.}
For each half, when a majority of the executions of
\cref{thm:all-LSS-algo} return the required cardinalities,
the medians equal these cardinalities at every sum.
When this holds for both halves, the chosen split gives
the largest possible cardinality of a subset of $I$ with sum $u$:
every such subset splits between the two halves, and conversely any
two subsets with these sums have union of sum $u$.
Induction on $|I|$ therefore proves the asserted output guarantee
when this majority condition holds at every recursive split.

\subparagraph*{Failure probability.}
Each execution has at least two input items, so its failure probability
is at most $2^{-10}$ by \cref{thm:all-LSS-algo}. For either half,
independence bounds the probability that at least $R/2$ executions
fail by
\[
 2^R(2^{-10})^{R/2}=2^{-4R}\leq\frac{1}{12n}.
\]
This bound holds conditioned on all preceding computations.
There are at most $2(n-1)$ halves on which we compute the medians.
A union bound gives total failure probability at most
$2(n-1)/(12n)<1/6$.

\subparagraph*{Running time.}
At a fixed depth of the recursion, let $h$ be the number of recursive
instances and let $n_i$ and $t_i$ be the size and target of instance
$i\in[h]$. Their index sets are disjoint and their targets sum to
at most $t$. Handling the base cases, computing the two \ATLCSS
arrays for each remaining instance, and scanning its $t_i+1$ possible
splits take a total time at that depth of
\[
 \Oh\!\left(
 \left(\sum_{i=1}^h n_i+\sum_{i=1}^h t_i^{3/2}\right)
 \operatorname{polylog}(n+t)\right)
 \leq\Oh\!\left((n+t^{3/2})
 \operatorname{polylog}(n+t)\right).
\]
Here $\sum_{i=1}^h n_i\leq n$ and
$\sum_{i=1}^h t_i^{3/2}\leq(\sum_{i=1}^h t_i)^{3/2}\leq t^{3/2}$.
There are at most $\lceil\log_2 n\rceil+1$ depths.
The $R=\Oh(\log(n+1))$ executions per half and the computation of
their medians contribute only to the displayed logarithmic factors.
Summing over all depths gives $\OL(n+t^{3/2})$ total time.
\end{proof}

For a nonempty $S\subseteq[n]$, define
\[
 q_S=\left\lfloor\frac{|t-\Sigma(S)|}{|S|}\right\rfloor,
 \qquad
 r_S=|t-\Sigma(S)|-|S|q_S,
 \qquad
 F_p(S)=(|S|-r_S)q_S^p+r_S(q_S+1)^p.
\]
For fixed $p\geq1$, evaluating and comparing values of $F_p$ take
constant time in our running-time bounds.
The following lemma shows that $F_p(S)$ is the minimum value of
$\norm{Z-X}_p^p$ subject to $\sum_{i\in S}x_i=t$.

\begin{lemma}\label{lem:pss-balanced-cost}
Fix $p\geq1$. For every nonempty $S\subseteq[n]$, the minimum of
$\norm{Z-X}_p^p$ over $X\in\Z^n$ satisfying
$\sum_{i\in S}x_i=t$ is $F_p(S)$.
Let $S^\star$ minimize $F_p(S)$ over nonempty
$S\subseteq[n]$, and choose any $R\subseteq S^\star$ with
$|R|=r_{S^\star}$. For an integer $z$, $\operatorname{sgn}(z)$ is
$1$ if $z>0$, $0$ if $z=0$, and $-1$ if $z<0$. The vector
$X=(x_1,\ldots,x_n)$ defined by
\[
 x_i=\begin{cases}
 a_i+\operatorname{sgn}(t-\Sigma(S^\star))(q_{S^\star}+1),&i\in R,\\
 a_i+\operatorname{sgn}(t-\Sigma(S^\star))q_{S^\star},
     &i\in S^\star\setminus R,\\
 a_i,&i\in[n]\setminus S^\star
 \end{cases}
\]
together with $S^\star$ is an optimal solution to $\PSS_p$, of cost
$F_p(S^\star)^{1/p}$. For a fixed sum $\Sigma(S)$, a subset of
largest cardinality minimizes $F_p(S)$.
\end{lemma}
\begin{proof}
Fix a nonempty $S\subseteq[n]$. Setting $x_i=a_i$ for
$i\in[n]\setminus S$ preserves
$\sum_{i\in S}x_i=t$ and cannot increase $\norm{Z-X}_p^p$.
Writing $b_i=x_i-a_i$ for $i\in S$, we minimize
$\sum_{i\in S}|b_i|^p$ over integers satisfying
$\sum_{i\in S}b_i=t-\Sigma(S)$.
For $i,j\in S$ with $b_i\geq b_j+2$, replacing $(b_i,b_j)$ by
$(b_i-1,b_j+1)$ does not increase $\sum_{i\in S}|b_i|^p$, by convexity of
$|x|^p$. Repeating these operations terminates because
$\sum_{i\in S}b_i^2$ strictly decreases. At termination, the integers
$b_i$ differ by at most one. Since their sum is $t-\Sigma(S)$, there are
$|S|-r_S$ copies of $\operatorname{sgn}(t-\Sigma(S))q_S$ and
$r_S$ copies of $\operatorname{sgn}(t-\Sigma(S))(q_S+1)$.
Thus the minimum of $\sum_{i\in S}|b_i|^p$ is $F_p(S)$.
The vector $X=(x_1,\ldots,x_n)$ satisfies
$\norm{Z-X}_p^p=F_p(S^\star)$, and the choice of $S^\star$
makes $X$ optimal for $\PSS_p$.

For nonempty $S_1,S_2\subseteq[n]$ with $\Sigma(S_1)=\Sigma(S_2)$
and $|S_1|\leq|S_2|$, choose integers $b_i$, for $i\in S_1$, satisfying
$\sum_{i\in S_1}b_i=t-\Sigma(S_1)$ and
$\sum_{i\in S_1}|b_i|^p=F_p(S_1)$.
Assign these integers to distinct indices of $S_2$, and assign zero
to the remaining indices. Denote the resulting integers by
$(c_i)_{i\in S_2}$. Then $\sum_{i\in S_2}c_i=t-\Sigma(S_2)$ and
$\sum_{i\in S_2}|c_i|^p=F_p(S_1)$.
Because $F_p(S_2)$ is the minimum value of $\norm{Z-X}_p^p$ among
vectors satisfying $\sum_{i\in S_2}x_i=t$, we have
$F_p(S_2)\leq F_p(S_1)$, proving the largest-cardinality assertion.
\end{proof}

We can now compute an optimal solution by choosing a sum and recovering
a largest subset with that sum.

\pssallnormsalgo*
\begin{proof}
We first describe the algorithm, prove its correctness, bound its failure
probability, and analyze its running time.

\subparagraph*{Description of the algorithm.}
Let $j_0\in[n]$ satisfy $a_{j_0}=\min_{i=1}^n a_i$.
If $a_{j_0}\geq t$, set $x_{j_0}=t$ and
$x_i=a_i$ for $i\in[n]\setminus\{j_0\}$, and return $(X,\{j_0\})$.
Otherwise, $a_{j_0}<t$, and we proceed as follows.
We solve \ATLCSS on $Z$ with sum cap $3t$ using
\cref{thm:all-LSS-algo} with $c=10$, and denote its output by $L$.
Let $D=\{s\in[3t]_0:L(s)>0\}$. For each $s\in D$, compute
\[
 \begin{aligned}
 q_s&=\left\lfloor\frac{|t-s|}{L(s)}\right\rfloor,
 \qquad r_s=|t-s|-L(s)q_s,\\
 C(s)&=(L(s)-r_s)q_s^p+r_s(q_s+1)^p.
 \end{aligned}
\]
Choose $t^\star\in D$ minimizing $C(t^\star)$, or return $\fail$ if $D$ is empty.
Apply the algorithm of \cref{lem:pss-largest-subset} to $Z$ with
target $t^\star$, using independent randomness. If this algorithm returns
$\fail$ or an empty subset, return $\fail$. Otherwise, let $S$ be
its output, construct $X$ according to \cref{lem:pss-balanced-cost},
and return $(X,S)$.

\subparagraph*{Correctness.}
If $a_{j_0}\geq t$, fix a nonempty $S\subseteq[n]$.
Then $\Sigma(S)\geq |S|a_{j_0}\geq t$. Applying Jensen's
inequality~\cite[Appendix~B]{roch2024modern} to the convex function
$x\mapsto |x|^p$, with weights $1-r_S/|S|$ and $r_S/|S|$, gives
\[
 \begin{aligned}
 \frac{F_p(S)}{|S|}
 &=\left(1-\frac{r_S}{|S|}\right)q_S^p
   +\frac{r_S}{|S|}(q_S+1)^p\\
 &\geq\left(q_S+\frac{r_S}{|S|}\right)^p
 =\left(\frac{\Sigma(S)-t}{|S|}\right)^p.
 \end{aligned}
\]
Moreover, since $t\geq0$ and $|S|\geq1$,
\[
 \frac{\Sigma(S)-t}{|S|}
 \geq a_{j_0}-\frac{t}{|S|}
 \geq a_{j_0}-t\geq0.
\]
Consequently,
\[
 F_p(S)\geq |S|\left(\frac{\Sigma(S)-t}{|S|}\right)^p
 \geq(a_{j_0}-t)^p=F_p(\{j_0\}).
\]
Thus the singleton $\{j_0\}$ is optimal.
It remains to consider $a_{j_0}<t$. We first justify the sum cap $3t$.

\begin{claim*}
There is a nonempty subset $S^\star$ minimizing $F_p$ with
$\Sigma(S^\star)\leq3t$.
\end{claim*}
\begin{claimproof}
Fix $S\subseteq[n]$ with $\Sigma(S)>3t$ and $|S|\geq2$.
Choose $j\in S$ with $a_j=\max_{i\in S}a_i$.
Choose nonnegative integers $b_i\in\{q_S,q_S+1\}$ for $i\in S$,
with exactly $r_S$ indices satisfying $b_i=q_S+1$, and with
$b_j=\max_{i\in S}b_i$. By the definitions of $q_S$, $r_S$, and $F_p$,
\[
 \sum_{i\in S}b_i=\Sigma(S)-t,
 \qquad \sum_{i\in S}b_i^p=F_p(S),
 \qquad b_j=\left\lceil\frac{\Sigma(S)-t}{|S|}\right\rceil.
\]
Let $B=\sum_{i\in S\setminus\{j\}}b_i$. Then
\[
 \begin{aligned}
 B&=\sum_{i\in S\setminus\{j\}}b_i
 =\Sigma(S)-t-\left\lceil\frac{\Sigma(S)-t}{|S|}\right\rceil\\
 &\geq\left\lfloor\frac{\Sigma(S)-t}{2}\right\rfloor\geq t.
 \end{aligned}
\]
If $\Sigma(S\setminus\{j\})\geq t$, then
$|t-\Sigma(S\setminus\{j\})|=\Sigma(S)-t-a_j\leq B$, since
$a_j\geq\lceil\Sigma(S)/|S|\rceil$.
Otherwise, $|t-\Sigma(S\setminus\{j\})|=t-\Sigma(S\setminus\{j\})\leq t\leq B$.
We can therefore choose integers $0\leq c_i\leq b_i$ for
$i\in S\setminus\{j\}$ with
$\sum_{i\in S\setminus\{j\}}c_i=|t-\Sigma(S\setminus\{j\})|$.
Setting $x_i=a_i+\operatorname{sgn}(t-\Sigma(S\setminus\{j\}))c_i$
for $i\in S\setminus\{j\}$ and $x_i=a_i$ for
$i\in[n]\setminus(S\setminus\{j\})$ gives a feasible
pair $(X,S\setminus\{j\})$. Since $b_j>0$,
\[
 F_p(S\setminus\{j\})\leq\sum_{i\in S\setminus\{j\}}c_i^p
 <\sum_{i\in S}b_i^p=F_p(S).
\]
For $i\in[n]$ with $a_i>3t$, the singleton $\{i\}$ satisfies
$F_p(\{i\})=(a_i-t)^p>(t-a_{j_0})^p=F_p(\{j_0\})$.
Thus no subset minimizing $F_p$ has sum greater than $3t$.
\end{claimproof}

Now suppose the algorithm of \cref{thm:all-LSS-algo}
returns the required cardinalities. Then $D$ consists exactly of the sums
at most $3t$ attainable by nonempty subsets, and $L(s)$ is the largest
cardinality of a subset of sum $s$ for every $s\in D$.
Let $S^\star$ minimize $F_p$ with $\Sigma(S^\star)\leq3t$, as guaranteed
by the claim, and write $s^\star=\Sigma(S^\star)$.
For each $s\in D$, $C(s)$ is the value of $F_p$ for any largest
subset of sum $s$. By \cref{lem:pss-balanced-cost},
$C(s^\star)\leq F_p(S^\star)$. Hence
\[
 C(t^\star)\leq C(s^\star)\leq F_p(S^\star).
\]
If the algorithm of \cref{lem:pss-largest-subset} also returns a largest
subset $S$ of sum $t^\star$, then $F_p(S)=C(t^\star)$.
Since $S^\star$ minimizes $F_p$ over all nonempty subsets, the
inequalities above must be equalities. Thus $S$ minimizes $F_p$, and
\cref{lem:pss-balanced-cost} guarantees that $(X,S)$ is feasible and
optimal for $\PSS_p$.

\subparagraph*{Failure probability.}
The execution of \cref{thm:all-LSS-algo} fails with probability at
most $n^{-10}$. Conditioned on its successful return and the selected
$t^\star$, the independent algorithm of \cref{lem:pss-largest-subset}
fails with probability at most $1/6$. The total failure probability is
therefore at most $n^{-10}+1/6<1/3$, since $n\geq2$.

\subparagraph*{Running time.}
The algorithm of \cref{thm:all-LSS-algo} with sum cap $3t$ and the
algorithm of \cref{lem:pss-largest-subset} with target $t^\star\leq3t$
each take $\OL(n+t^{3/2})$ time. Computing $C(s)$ for $s\in D$ and writing
$X$ and $S$ take $\Oh(n+t)$ additional time.
\end{proof}

\subsection{Faster algorithm for Projection Subset Sum on the
\texorpdfstring{$\ell_1$}{ell1} norm}

For $\PSS_1$, it is in fact enough to find a sum closest to $t$ among sums of
nonempty subsets. This allows us to solve $\PSS_1$ in near-linear pseudopolynomial
time. First, we recover a subset with that sum using Bringmann's
algorithm~\cite{bringmann2017near}.

\begin{lemma}\label{lem:pss-subset-search}
Given nonnegative integers $a_1,\ldots,a_n$, an attainable integer
sum $t\geq0$, there is a randomized algorithm
that returns a subset $S\subseteq[n]$ with $\Sigma(S)=t$, or $\fail$.
It returns $\fail$ with probability at most $1/6$ and runs in time
$\OL(n+t)$.
\end{lemma}
\begin{proof}
We first describe the algorithm, prove its correctness, bound its failure
probability, and analyze its running time.

\subparagraph*{Description of the algorithm.}
We use Bringmann's algorithm~\cite[Theorem~1.1]{bringmann2017near}:
on positive items with sum cap $u>0$ and error parameter $\eta\in(0,1)$,
it outputs only attainable sums at most $u$ and omits any fixed
attainable sum with probability at most $\eta$, in time
$\Oh((|I|+u)\operatorname{polylog}(|I|+u+1/\eta))$
for an input indexed by $I$.

Set $\eta=1/(12n)$ and start with $I=[n]$ and $u=t$.
For $u=0$, return the empty set.
If $I=\{i\}$ and $u>0$, we return $\{i\}$ when $a_i=u$, and $\fail$
otherwise. For $|I|\geq2$ and $u>0$, we split
$I=I_1\uplus I_2$ with $\bigl||I_1|-|I_2|\bigr|\leq1$.
For each $r\in\{1,2\}$, apply Bringmann's algorithm to
$(a_i)_{i\in I_r}$ with sum cap $u$ and error parameter $\eta$,
using randomness independent of all previous executions.
Discard items of value zero or greater than $u$ before this execution,
and let $\mathcal A_r$ be its output together with $0$.
If no item remains, set $\mathcal A_r=\{0\}$; if one item remains,
compute its subset sums directly.
We find $s\in[u]_0$ with $s\in\mathcal A_1$ and
$u-s\in\mathcal A_2$, returning $\fail$ if none exists.
Apply the same procedure to $(I_1,s)$ and $(I_2,u-s)$.
If either invocation returns $\fail$, return $\fail$; otherwise,
return the union of the two subsets it outputs.

\subparagraph*{Correctness.}
Since $\mathcal A_r$ contains only sums of subsets of $I_r$,
every recursive target is attained by a subset of its recursive
index set. Induction on $|I|$ shows that the returned set has sum
$u$, since $I_1$ and $I_2$ are disjoint.
\subparagraph*{Failure probability.}
Conditioned on the recursion history at $(I,u)$, fix a subset
$S\subseteq I$ of sum $u$. Bringmann's algorithm reports both
$\Sigma(S\cap I_1)$ in $\mathcal A_1$ and
$\Sigma(S\cap I_2)$ in $\mathcal A_2$ with probability at least
$1-2\eta$, so the split search succeeds with at least this probability.
There are at most $n-1$ recursive splits. A union bound over their
split searches gives failure probability at most $2(n-1)\eta<1/6$.
\subparagraph*{Running time.}
At each depth of the recursion, the recursive index sets are disjoint
and their targets sum to at most $t$. Executing Bringmann's algorithm
on both halves of each index set and scanning the possible split sums take
$\Oh((n+t)\operatorname{polylog}(n+t+1/\eta))$ time per depth.
There are $\Oh(\log(n+1))$ depths. Summing over the depths with
$1/\eta=12n$ gives $\OL(n+t)$ total time.
\end{proof}

We now use \cref{lem:pss-subset-search} to output the modified vector
and a set whose modified values sum to $t$.

\begin{theorem}\label{thm:pss-l1-algo}
Given nonnegative integers $Z=(a_1,\ldots,a_n)$ and an integer $t\geq0$,
there is a randomized algorithm that outputs an optimal solution
$(X,S)$ to $\PSS_1$ in time $\OL(n+t)$, with failure probability
at most $1/3$.
\end{theorem}
\begin{proof}
We first describe the algorithm, prove its correctness, bound its failure
probability, and analyze its running time.

\subparagraph*{Description of the algorithm.}
Let $j_0\in[n]$ satisfy $a_{j_0}=\min_{i=1}^n a_i$.
If $a_{j_0}\geq t$, set $x_{j_0}=t$ and
$x_i=a_i$ for $i\in[n]\setminus\{j_0\}$, and return $(X,\{j_0\})$.

For $a_{j_0}<t$, proceed as follows.
Apply Bringmann's algorithm~\cite[Theorem~1.1]{bringmann2017near}
to the items of $Z$ with values in $[2t]$, with sum cap $2t$ and
error parameter $1/6$. If at most one such item exists, compute its
subset sums directly. Let $\mathcal A$ be the resulting set of sums
together with $a_{j_0}$, delete $0$ if $a_{j_0}>0$, and choose
$t^\star\in\mathcal A$ minimizing $|t-t^\star|$.

If $t^\star=0$, set $S=\{j_0\}$; otherwise, apply the algorithm of
\cref{lem:pss-subset-search} to $Z$ with target $t^\star$, using randomness
independent of the preceding Bringmann computation.
If this algorithm returns $\fail$, return $\fail$; otherwise, let $S$
be its output.
Choose any $j\in S$, set $x_j=a_j+t-t^\star$ and
$x_i=a_i$ for $i\in[n]\setminus\{j\}$, and return $(X,S)$.

\subparagraph*{Correctness.}
For $t>0$, any feasible vector $X'=(x'_1,\ldots,x'_n)$ has a
nonempty $S'\subseteq[n]$ with $\sum_{i\in S'}x'_i=t$, so
\[
 \norm{Z-X'}_1
 \geq\left|\sum_{i\in S'}(a_i-x'_i)\right|
 =|\Sigma(S')-t|.
\]
If $a_{j_0}\geq t$, every nonempty $S'$ has
$\Sigma(S')\geq a_{j_0}$, so $a_{j_0}-t$ is a lower bound on the optimal cost.
The returned pair attains this lower bound.

For $a_{j_0}<t$, fix a sum $s^\star$ of a nonempty subset
minimizing $|t-s^\star|$, independently of the algorithm's randomness.
Since $\Sigma(\{j_0\})=a_{j_0}$ and $t-a_{j_0}\leq t$, whereas
$|t-s|>t$ for $s>2t$, we have $s^\star\leq2t$.
The set $\mathcal A$ is nonempty and contains only sums attained by
nonempty subsets. If $s^\star\in\mathcal A$, the chosen $t^\star$ is closest
to $t$ among sums of nonempty subsets.
When the algorithm of \cref{lem:pss-subset-search} returns a subset
$S$ of sum $t^\star$, or $t^\star=0$ and $S=\{j_0\}$, the returned vector and $S$ are
feasible and have cost $|t-t^\star|$. Whenever $t^\star$ is closest to $t$ among
sums of nonempty subsets, this cost attains the lower bound on
$\norm{Z-X'}_1$ for every feasible vector $X'$.

\subparagraph*{Failure probability.}
The probability that $s^\star$ is omitted is at most $1/6$.
For $t^\star>0$, conditioned on the selected $t^\star$, the independent algorithm
of \cref{lem:pss-subset-search} returns $\fail$ with probability at
most $1/6$. Hence an optimal pair $(X,S)$ is returned with probability
at least $2/3$.

\subparagraph*{Running time.}
Since $t^\star\leq2t$, Bringmann's algorithm with sum cap $2t$ and the
algorithm of \cref{lem:pss-subset-search} with target $t^\star$ each take
$\OL(n+t)$ time, and the
scan of $\mathcal A$ and construction of $X$ and $S$ take $\Oh(n+t)$ time.
\end{proof}
 \bibliographystyle{plainurl}
\bibliography{refs}
\appendix
\section{Deferred Lemmas and Proofs} \label{sec:deferred-proofs}

\subsection{Fast convolution of sets of pairs} \label{sec:pair-sumset}

For finite sets $P, Q \subseteq \mathbb{Z}^2$, we show how to compute their convolution $P \oplus Q$ efficiently.

\begin{lemma}\label{lem:pair-sumset}
Let $P,Q\subseteq\mathbb \{-M_1, \ldots, M_1\} \times \{-M_2, \ldots, M_2\}$, then $P\oplus Q$ can be computed in time $\OL(M_1M_2)$.
\end{lemma}
\begin{proof}
Set $b=4M_1+1$ and represent a pair $(x,y)$ in either input set by
the exponent $(x+M_1)+b(y+M_2)$. We form a polynomial for each set
whose coefficients indicate the represented pairs, then compute their
product by exact integer convolution (for example, using fast Fourier transform; see \cite{cormen2025algorithmen}). The exponent contributed by a
pair $(x,y)\in P\oplus Q$ is
$(x+2M_1)+b(y+2M_2)$. Since $0\leq x+2M_1\leq4M_1<b$, there is no
carry between the coordinates. Thus a coefficient of the product is
positive exactly when its exponent represents a pair in $P\oplus Q$.
The polynomials and their product have degree $\Oh(M_1M_2)$, and each
coefficient of the product is at most $|P||Q|=\Oh(M_1^2M_2^2)$.
Exact fast convolution and a scan of the coefficients therefore take
$\OL(M_1M_2)$ time.
\end{proof}

\subsection{Proof of Lemma~\ref{lem:all-targets-bounded-cardinality}}\label{sec:proof-of-all-targets-bounded-cardinality}
We recall \cref{lem:all-targets-bounded-cardinality} from \ref{sec:overview} and prove it.
\AllTCardBound*
\begin{proof}    
The proof follows the generating-function approach of Jin and Wu~\cite{JinWu19}, with an
additional variable recording the cardinality. The algorithm computes
coefficients modulo a random prime. A zero integer coefficient is never
reported as nonzero, but a positive coefficient may vanish modulo the
chosen prime. We bound this one-sided error below.

Since the $a_i$ are nonnegative, an index $i$ with $a_i>t$ cannot belong to any
subset $S$ with $\Sigma(S)\le t$, and we henceforth disregard such indices.
Let $z:=|\{i\in[n]:a_i=0\}|$.

For a bivariate formal power series $F(x,y)$, let
\[
    [x^s y^r]F(x,y)
\]
denote the coefficient of the monomial $x^s y^r$.  We also define its
$(t,k)$-truncation by
\[
    \operatorname{tr}_{t,k}(F)
    :=
    \sum_{s=0}^{t}\sum_{r=0}^{k}
       [x^s y^r]F(x,y)\,x^s y^r.
\]
Thus, after every polynomial operation, terms of $x$-degree larger than $t$
or $y$-degree larger than $k$ may be discarded.

Consider the following bivariate generating function defined by
\[
    A(x,y)
    :=
    \prod_{i=1}^{n}\bigl(1+yx^{a_i}\bigr).
\]
The subset sum instance has a solution for sum $s$ of cardinality $k$ if and only if 
$[x^s y^r]A(x,y) \neq 0$. Formally, we have for every $0\le s\le t$ and $0\le r\le k$,
\[
    [x^s y^r]A(x,y)
    =
    \bigl|
        \{S\subseteq[n]:
          \Sigma(S)=s,\ |S|=r\}
    \bigr|.
\]

We first explain how to compute these coefficients modulo a prime.  Let
$p>\max\{t,k\}$ be a prime, and let $\mathbb F_p$ denote the residue ring modulo $p$. 
All computations in the remainder of this paragraph are over
$\mathbb F_p$. Let
\[
    B(x,y):=\ln\!\left(\prod_{i : a_i > 0}(1+yx^{a_i})\right) \in \mathbb{Q}[[x, y]].
\]

\begin{proposition}\label{prop: log_computation}
    For prime $p > \max\{k, t\}$, in $\widetilde O(n+t)$ time one can compute 
    $\operatorname{tr}_{t,k}(B)$.
\end{proposition}
\begin{proof}
    The zero-valued items contribute the factor $(1+y)^z$ to $A(x,y)$.
    Over the rational formal power series, using
    \[
        \ln(1+v)=\sum_{j\ge1}\frac{(-1)^{j-1}}{j}v^j,
    \]
    we have
    \[
    \begin{aligned}
        B(x,y)
        &=
        \sum_{i:a_i>0}\ln\bigl(1+yx^{a_i}\bigr)\\
        &=
        \sum_{i:a_i>0}
        \sum_{j\ge1}
           \frac{(-1)^{j-1}}{j}
           y^j x^{j a_i}.
    \end{aligned}
    \]
    For every $a\in[t]$, let
    \[
        m_a:=|\{i\in[n]:a_i=a\}|
    \]
    be the multiplicity of $a$ among the input integers.  Truncating (2) gives
    \[
        \operatorname{tr}_{t,k}(B)
        =
        \sum_{a=1}^{t}
           \sum_{j=1}^{\min\{k,\lfloor t/a\rfloor\}}
           m_a\frac{(-1)^{j-1}}{j}x^{aj}y^j.
    \]
    Every denominator occurring in (3) satisfies $1\le j\le t$, and is therefore
    invertible modulo $p$.  Hence (3) also unambiguously defines
    $\operatorname{tr}_{t,k}(B)$ over $\mathbb F_p$.
    
    The multiplicities $m_1,\ldots,m_t$ can be computed in $O(n+t)$ time.
    Moreover, the number of pairs $(a,j)$ occurring in (3) is at most
    \[
        \sum_{a=1}^{t}\left\lfloor\frac{t}{a}\right\rfloor
        = O(t\ln t).
    \]
    Thus all coefficients of $\operatorname{tr}_{t,k}(B)$ can be computed in
    $\widetilde O(n+t)$ time.
\end{proof}

It remains to recover
\[
    \operatorname{tr}_{t,k}(A)
    =
    \operatorname{tr}_{t,k}\bigl((1+y)^z\exp(B)\bigr).
\]
We use the divide-and-conquer computation of the exponential formal power
series of Jin and Wu~\cite[Lemma~2]{JinWu19}; for completeness, we verify
that adding the cardinality variable increases its running time by only a
factor $k$.
\begin{proposition} \label{prop: Brent_bivariate}
    Given a polynomial $f(x, y) \in x \mathbb{F}_p[x, y]$ of degree at most $t$ $(< p)$, one can
    compute $\operatorname{tr}_{t,k}\bigl((1+y)^z\exp(f)\bigr)$ in time $\Tilde{\mathcal{O}}(tk)$.
\end{proposition}
\begin{proof}
    We write
    \[
        f(x,y)=\sum_{d=1}^{t} f_d(y)x^d
    \]
    for the polynomial $\operatorname{tr}_{t,k}(B)$, where every $f_d(y)$ has
    degree at most $k$, and we write
    \[
        g(x,y)
        =
        \operatorname{tr}_{t,k}\bigl((1+y)^z\exp(f(x,y))\bigr)
        =
        \sum_{i=0}^{t}g_i(y)x^i,
    \]
    where every $g_i(y)$ is taken modulo $y^{k+1}$. Since $f$ has zero
    $x$-constant term, $g_0(y)=(1+y)^z\bmod y^{k+1}$. Its coefficients
    $\binom{z}{j}$, for $0\leq j\leq k$, are computed in $O(k)$ field
    operations using $p>k$. The identity
    \[
        \frac{\partial g}{\partial x}
        =
        g\frac{\partial f}{\partial x}
    \]
    implies, by comparing coefficients of $x^{i-1}$, that for every
    $1\le i\le t$,
    \[
        i\,g_i(y)
        =
        \sum_{j=0}^{i-1}
           (i-j)f_{i-j}(y)g_j(y)
           \pmod{y^{k+1}}.
    \]
    Since $p>t$, the integer $i$ is invertible in $\mathbb F_p$ for every
    $i\in[t]$, so (4), together with $g_0(y)=(1+y)^z\bmod y^{k+1}$, uniquely determines
    $g_1(y),\ldots,g_t(y)$.
    
    We compute the recurrence (4) by divide and conquer on the index interval
    $[0,t]$, exactly as in the univariate algorithm of
    Jin and Wu~\cite[Lemma~2]{JinWu19}.  We consider one recursive interval
    $[\ell,r]$, and let
    \[
        q:=\left\lfloor\frac{\ell+r}{2}\right\rfloor,
    \]
    and we suppose that $g_\ell(y),\ldots,g_q(y)$ have already been computed.
    Their total contribution to the right-hand coefficients
    $g_{q+1}(y),\ldots,g_r(y)$ in (4) is obtained from the product
    \[
        \left(
           \sum_{d=1}^{r-\ell} d\,f_d(y)z^d
        \right)
        \left(
           \sum_{j=\ell}^{q}g_j(y)z^{j-\ell}
        \right).
    \]
    Indeed, for every $q<i\le r$, the coefficient of $z^{i-\ell}$ in (5) is
    \[
        \sum_{j=\ell}^{q}
           (i-j)f_{i-j}(y)g_j(y),
    \]
    which is exactly the part of the right-hand side of (4) contributed by
    indices $j\in[\ell,q]$.
    
    We next bound the time needed to compute (5).  We suppose the length of the
    recursive interval is
    \[
        L:=r-\ell+1.
    \]
    Both factors in (5) have $z$-degree $O(L)$ and $y$-degree at most $k$.
    Their product, truncated modulo $y^{k+1}$, can be computed in
    $\widetilde O(Lk)$ field operations.  To see this using only univariate
    polynomial multiplication, substitute
    \[
        z\longmapsto Z^{2k+1},
        \qquad
        y\longmapsto Z.
    \]
    Before truncation, the product of the two factors has $y$-degree at most
    $2k$.  Therefore different monomials $z^a y^b$, with
    $0\le b\le 2k$, are mapped to different powers
    \[
        Z^{(2k+1)a+b}.
    \]
    Hence one multiplication of two univariate polynomials of degree
    $O(Lk)$ suffices to compute (5), after which the coefficients of
    $y$-degree larger than $k$ are discarded.  Fast univariate polynomial
    multiplication takes $\widetilde O(Lk)$ time
    (cf.\ \cite[Lemma~1]{JinWu19}).
    
    At every fixed depth of the divide-and-conquer recursion, the recursive
    intervals are disjoint, and their lengths sum to $O(t)$.  Consequently, the
    total cost of all polynomial multiplications at one recursion depth is
    $\widetilde O(tk)$.  There are $O(\ln t)$ recursion depths, and all other
    operations require at most the same amount of time.  Thus
    \[
        \operatorname{tr}_{t,k}\bigl((1+y)^z\exp(f)\bigr)
    \]
    is computed in
    \[
        \widetilde O(tk)
    \]
    time.
\end{proof}

\begin{proposition}\label{prop: computing_A_mod_p}
    For prime $p > \max\{k, t\}$, one can compute $[x^sy^r]A(x, y) \mod p$ for all $0 \leq s \leq t$
    and $0 \leq r \leq k$ in time $\Tilde{\mathcal{O}}(n + tk)$ time.
\end{proposition}
\begin{proof}
    Firstly, we use \cref{prop: log_computation} to compute 
    $[x^sy^r]B(x, y) \mod p$ for all $0 \leq s \leq t$
    and $0 \leq r \leq k$ in time $\Tilde{\mathcal{O}}(n + t)$ time, and then
    we use \cref{prop: Brent_bivariate} to compute 
    $\operatorname{tr}_{t,k}\bigl((1+y)^z\exp(B)\bigr)$ in $\widetilde O(tk)$ time.
    And, over the rational formal power series, we have
    \[
        A(x,y) = (1+y)^z\exp(B(x,y)).
    \]
    This completes the proof.
\end{proof}

It remains only to ensure that reduction modulo $p$ does not turn a
positive coefficient in (1) into zero.  We use the same random-prime
argument as Jin and Wu~\cite[Lemma~3,Theorem~6]{JinWu19}, with a union
bound over all pairs $(s,r)$ we have the following

\begin{corollary}\label{cor: prime_not_divides}
    Let $M = n+(k+1)t + \ceil{1/\eta}$ where $\eta \in \{0, 1\}$ 
    and $p$ be a uniform random prime from $[M^4, 2M^4]$. 
    Then
    \[
    \Pr\!\left[
       p \mid [x^sy^r]A(x, y)
       \text{ for some }(s,r)\text{ with } [x^sy^r]A(x, y) \ne 0
    \right] \leq \eta.
\]
\end{corollary}

By sampling and using Miller-Rabin primality testing~\cite{cormen2025algorithmen}, 
we can pick a uniform random prime $p$ from interval $[M^4, 2M^4]$ 
in $\operatorname{poly}\log(M)$ time with $\mathcal{O}(M^{-1})$ failure probability. 
Then the lemma immediately follows from~\cref{cor: prime_not_divides} and
\cref{prop: computing_A_mod_p}.
\end{proof}

\end{document}